\documentclass[12pt]{article}
\usepackage{graphicx}
\usepackage{epsfig}
\usepackage{rotating}
\usepackage{amsmath}
\usepackage{amsfonts}
\usepackage{amssymb}
\usepackage{url}
\usepackage{hyperref}
\usepackage[footnotesize]{caption2}
\newcommand{\ba}{\begin{array}{c}}
\newcommand{\ea}{\end{array}}
\newcommand{\baz}{\begin{array}{cc}}
\def\ba{\begin{eqnarray}}
\def\ea{\end{eqnarray}}
\def\br{\begin{array}}
\def\er{\end{array}}
\def\be{\begin{equation}}
\def\ee{\end{equation}}

\def\d{\delta}

\def\da1{d\alpha_1\over dt}
\def\da2{d\alpha_2\over dt}

\def\cj{{\cal J}}

\newcommand{\bea}{\begin{eqnarray}}
\newcommand{\eea}{\end{eqnarray}}
\newcommand{\bmt}{\begin{pmatrix}}
\newcommand{\emt}{\end{pmatrix}}
\begin{document}

\par\noindent{ \large\bf { Singlet Fermion Assisted Dominant Seesaw with
    Lepton Flavor and Number Violations and Leptogenesis}}\\
 
\begin{center}
\vskip .15in
{\bf M. K. Parida ${}^{*}$, Bidyut Prava Nayak}\\

{\sl${}^*${{\em Centre of Excellence in Theoretical and Mathematical
~Sciences\\
{\rm SOA} University,
 Khandagiri Square, Bhubaneswar  751030, India}}}\\ 
\end{center}

\begin{abstract}
In a recent review  Mohapatra has discussed how type-I seesaw
mechanism suppressed by fine tuning or specific textures of associated
fermion mass matrices can form the basis of neutrino masses in TeV
scale $W_R$ boson models. In this paper we review recent works in
another class of theories 
where the added presence of gauge singlet fermions render the type-I seesaw contribution
vanishing  but extended seesaw dominant. In this case the light
neutrino mass formula is the same as the classic inverse seesaw
derived earlier but the singlet fermion masses are governed by a
separate type-I seesaw like formula. 
 Embeddings of this mechanism in
 supersymmetric as well as non-supersymmetric SO(10) with low or
 intermediate masses of
 $W_R$ or $Z_R$ bosons are discussed. We also discuss how this
 cancellation criterion has led to a new mechanism of type-II seesaw dominance which  permits $U(1)_{B-L}$ breaking scale much smaller than the left-handed triplet mass. Out of a number of new observable
predictions, the most visible one in both cases are charged lepton
flavor violating decays accessible to ongoing searches and  dominant contribution to double
beta decay mediated by the light gauge singlet fermions in the $W_L-W_L$
channel. These seesaw dominance mechanisms  are
applicable in the extensions of the SM and high, intermediate, or low
scale left-right gauge theories  with or without their SO(10)
origin. Recent applications of this mechanism
covering dark matter and leptogenesis are reviewed.
 Emergence of other dominant seesaw mechanisms are also briefly
pointed out.
     
\end{abstract}

\section{INTRODUCTION}\label{Sec.1}
 Although the standard model (SM) of strong, weak, and electromagnetic
 interactions has enjoyed tremendous success through numerous
 experimental tests, it has outstanding failures in three notable areas : neutrino masses, baryon asymmetry of the
universe, and dark matter. Two of the other conceptual and theoretical
difficulties are that the SM can not explain disparate values of its
gauge couplings, nor can it account for the monopoly of parity violation in weak
interaction leaving out strong, electromagnetic, and gravitational
 interactions. In hitherto unverified extensions of the SM in the scalar, fermion, or gauge sectors, however, there are different theories for
neutrino masses \cite{Rev:06-11,Rev:Alta},
which mainly exploit 
 various see-saw mechanisms
\cite{RNM:rev16,Valle:rev16,type-I,type-I-Valle:1980,type-II,rnm-gs:1981,Schecter-Valle-1:1982,Schecter-Valle-2:1982,Inverse,Inv:RNM:1986,Inv:Albright:1986,Inv:Witten:1986,Inv:Nandi:1986,Inv:Valle:1991,Inv:Wyler:LRS,Bernabeu:1987,Inv:Ma:1987,type-III,Extended:2000,Babu-Pati-Wil:2000,Linear:Barr:2004,Albright-Barr:2004,Linear:Malinsky-Valle:2005,Lindner:schmidt-smirnov:2005,Linear:Fukuyama:2007,Kang-Kim:2006,Majee-mkp:PLB:2007,Mitra-gs-vissani:2012}.\\

The most popular method of neutrino mass generation has been through
type-I or canonical seesaw mechanism \cite {type-I} which was noted  
 to apply in the simplest extension of the SM through
right-handed (RH) neutrinos encompassing family mixings
\cite{type-I-Valle:1980}.  
Most of the problems of the SM have potentially 
 satisfactory solutions in the minimal left-right symmetric (LRS)
 \cite{rnm-gs:1981,ps,rnmpati}  grand
 unified theory based on $SO(10)$
 \cite{georgi:1974,cmp1:1984,cmp2:1984,cmgmp:1985,rnm-mkp:1993,lmpr:1995,mkp-S4:2008,Kadastic:dm,mkp-pks-kb:dm:2010,Chu-Smirnov:2016}. Although
 the neutrino
 masses measured by the oscillation data \cite{nudata} are most simply accommodated in
 $SO(10)$ if both the left-handed (LH) and the right-handed (RH) neutrinos 
  are Majorana fermions, alternative interpretations in favour of Dirac neutrino masses
 within the standard model paradigm have been also advanced 
\cite{Valle:rev16,Valle:2016,Gluza:2016}. Currently a number of 
experiments \cite{bbexpt1,bbexpt2,bbexpt3,bbexpt4} on neutrinoless double beta ($0\nu\beta\beta$)
decay are in progress to resolve the issue on the Dirac
or Majorana
nature of neutrinos
\cite{rnm-gs:1981,Schecter-Valle-1:1982,Schecter-Valle-2:1982}.\\

 Another set of physical
processes under active experimental investigation  are charged lepton flavor
violating (LFV) decays,  $\tau\to e\gamma$, $\tau\to \mu\gamma$ ,
$\mu\to e\gamma$, $\mu\to e{\bar e}e$
 where the minimally extended SM embracing small
neutrino masses and GIM mechanism predicts branching ratios many orders smaller
than their current experimental limits
\cite{lfvexpt1,lfvexpt2,lfvexpt:BaBar,lfvexpt3}. However
supersymmetric theories possess high potential to explain LFV decays
closer to the current limits \cite{Babu:NOON2004,Petcov:2005,Morisi-Valle:2012,Deppisch:2012,de Gouvea:2013,Boucenna-Morisi-Valle:rev14,Heeck:2016,Riazuddin:1981}.
 
A special feature of left-right gauge theories \cite{ps,rnmpati} and SO(10) 
grand
unified theory (GUT)
\cite{georgi:1974,cmp1:1984,cmp2:1984,cmgmp:1985,rnm-mkp:1993,lmpr:1995,mkp-S4:2008}
is that the canonical seesaw formula 
\cite{type-I,type-I-Valle:1980}
for Majorana neutrino masses 
is invariably accompanied by the type-II seesaw formula
\cite{type-II,rnm-gs:1981,Schecter-Valle-1:1982}
\begin{equation}                     
          M_\nu=m^{II}_\nu +m^{I}_\nu = fv_L-M_D\frac{1}{M_N}M_D^T.  
\label{comb}
\end{equation}
The parameters entering into this hybrid seesaw formula have
fundamentally appealing  interpretations in  Pati-Salam model \cite{ps} or
SO(10) GUT.  
In eq.(\ref{comb}) $M_D(M_N)$ is Dirac (RH-Majorana) neutrino mass, $v_L \sim \lambda \frac{v_{wk}^2V_R}{M_{\Delta_L}^2}$ is the
induced vacuum expectation value (VEV) of the LH triplet
$\Delta_L$, $V_R=SU(2)_R\times U(1)_{B-L}$ breaking VEV of the RH
triplet $\Delta_R$, and $f$ is the Yukawa coupling of the
triplets $\subset {\overline {126}}_H$ of SO(10). The same Yukawa coupling $f$ also
defines the RH neutrino mass $M_N=fV_R$. Normally, because of the underlying
quark-lepton symmetry in SO(10) or Pati-Salam model, $M_D$ is of the
same order as $M_u$, the up-quark mass matrix, that drives the
canonical seesaw
scale to be large, $M_N\sim 10^{14}$ GeV. In the LR theory based upon
$SU(2)_L\times SU(2)_R\times U(1)_{B-L}\times SU(3)_C(\equiv G_{2213})$, $M_D\sim
M_l=$ charged lepton mass matrix. The neutrino oscillation data then
pushes this seesaw scale to  $M_N\sim 10^{10}$ GeV.  Similarly the
type-II seesaw scale is also around this mass.
With such high seesaw scales in non-supersymmetric (non-SUSY) SO(10)
model or LRS theory, there is no possibility of direct experimental
verification of the seesaw mechanism or the associated $W_R$ boson mass in
near future. Likewise, the predicted LFV decay rates are far below the experimental limits.
 
On the other hand, if experimental investigations at the Large Hadron
Collider (LHC) \cite{CMS:2012} are to confirm \cite{KS:1982}  TeV
scale $W_R$ \cite{RNM:rev16,Chien-psb-rnm:2013,mkp-sahoo:Proc:2014,mkp-sahoo:NP:2015,psb-rnm:so10:2015}
or $Z_R$ boson production \cite{bpn-mkp:2015,bpn-mkp:disp:2015}
accompanied by RH Majorana neutrinos in the like-sign dilepton
 channel with jets, $pp\to l^{\pm}l^{\pm} jj X$, this  should be
also consistent with the neutrino oscillation data \cite{nudata} through seesaw
mechanisms. This is possible if the relevant seesaw scales are brought
down to $\sim (1-10)$ TeV. \footnote{In the RH neutrino extension of the SM,
  explanation of neutrino oscillation data with 
   baryon asymmetry of the universe  has been addressed  for ${\cal
    O} (1-10)$ GeV type-I seesaw scale \cite{Shaposhnikov:2012}. Interesting
connections with
double beta decay and dilepton
  production with displaced vertices at LHC have been also discussed 
  with such low canonical seesaw scales \cite{Helo} and the
  models need  fine-tuning of the $\nu-N-\phi$ Yukawa coupling $y\sim
  10^{-8}$ which is about $3$ orders smaller than the electron Yukawa
    coupling. Neutrino  mass generation with standard model
    paradigm and their interesting applications have been discussed in a recent review ref.\cite{Valle:rev16}}. Observable displaced vertices within SM extension has been also discussed in \cite{Antusch:2016} and references therein.

The scope and applications of type-I seesaw to TeV scale $W_R$ boson
models have been discussed in the recent interesting review
\cite{RNM:rev16}. In such models D-Parity is at first broken at high
scale that makes the left-handed triplet much heavier than the $W_R-$
mass, but keeps the $G_{2213}(g_{2L}\neq g_{2R})$ unbroken down to 
much lower scale \cite{cmp1:1984,cmp2:1984,cmgmp:1985}. This causes the type-II
seesaw contribution of the hybrid seesaw formula of eq. (\ref{comb}) to be 
severely damped out in the LHC scale $W_R$ models where type-I seesaw
dominates. But because $M_N$ is also at the TeV scale, the predicted
type-I seesaw contribution to light neutrino mass turns out to be $10^6-10^{11}$
times larger than the experimental values unless it is adequately suppressed
while maintaining its dominance over type-II seesaw. Such suppressions
have been made possible in two ways:(i) using fine tuned values of
the Dirac neutrino mass matrices $M_D$ \cite{RNM:rev16,Helo,Antusch} ,(ii) introducing specific
textures to the fermion mass matrices $M_D$ and/or $M_N$
\cite{Petcov:2005,Chien-psb-rnm:2013,Dutta-RNM:2003,Pilaftsis-Underwood:2004,Kersten-Smirnov:2007,Lee-psb-rnm:2013,Barr-texture-tIII-lepto,Barr-Dirac-texture-so10,Hybrid-Inv-1,Hybrid-Inv-2,Hybrid-Inv-3,Valle:LinInv:A4:2009,Joshipura-hybrid,Deppisch:hybrid:2014}.

Possible presence of
 specific textures of constituent matrices in the context of inverse, linear
or type-I seesaw models have been also explored \cite{Adhikary:A4:2006,Mainak:2012,PRoy:2013,Mainak:2014,Ghosal:JHEP:2016,Ghosal:NP:2016,Ghosal:PLB:2016,ARC:A4S3:2016}.
More recently the hybrid seesaw ansatz for matter parity conserving
SO(10) has been applied to explain neutrino masses, dark matter, and
baryon asymmetrty of the universe without invoking any texture or intermediate
scale in the non-supersymmetric SO(10) ramework \cite{mkp-hybrid}.
  
Even without going beyond the SM paradigm and treating the added RH neutrinos
in type-I seesaw as gauge singlet fermions at $\sim $ GeV scale, rich structure of new physics
has been predicted including neutrino masses, dark matter, and baryon
asymmetry of the universe. The fine-tuned value of the associated
Dirac neutrino Yukawa coupling in these models is $y\sim 10^{-7}$ \cite{Shaposhnikov:2012}.

There are physical situations where type-II seesaw dominance, rather
than type-I seesaw or inverse seesaw, is
desirable
\cite{type-I-Valle:1980,type-II,rnm-gs:1981,Schecter-Valle-1:1982,
Schecter-Valle-2:1982,bpn-mkp:2015,Bajc-gs-vissani:2003,Goh-rnm-ng:2003,typeIISO10,HUM:2004,
Goh-rnm-nasri:2004,Dutta-mimura-rnm:2009,
rnmmkp11,gs:2015,Sruba:2015}. 

In
the minimal case, being a mechanism driven 
by intermediate scale mass of LH triplet, type-II seesaw
 may not be directly verifiable; nevertheless it can be clearly
applicable to TeV scale $Z_R$ models in non-SUSY SO(10) to account for neutrino
masses \cite{bpn-mkp:2015} provided type-I contribution is adequately suppressed. However,  as in the fine-tuning of Dirac neutrino mass in the 
 type-I seesaw case in LR models, the induced VEV needed for type-II seesaw can also be fine tuned using more than one electroweak bi-doublets reducing the triplet mass to lower scales accessible to accelerator tests.
Looking to the eq.(\ref{comb}) and the
structure of the
induced VEV $v_L$, the  most convenient method of suppressing type-I
seesaw with respect to type-II seesaw is to make the type-I seesaw scale
$M_N=fV_R$ larger and the triplet mass much smaller, $M_{\Delta_L}<< M_N$. This requires the $SU(2)_R\times   
U(1)_{B_L}$ breaking scale or $M_{W_R} >> M_{\Delta_L}$. SUSY and
non-SUSY SO(10) models have been constructed with this possibility
and also in the case of split-SUSY \cite{Goh-rnm-nasri:2004,rnmmkp11}
where $M_{W_R} \simeq 10^{17}$ GeV. Obviously such models have no
relevance in the context of TeV scale $W_R$ or $Z_R$ bosons accessible
to LHC searches.      

Whereas the pristine type-I or type-II seesaw  are essentially high scale
formulas inaccessible for direct verification 
and need fine tuning or textures to bring them down to the TeV scale, the well
 known  classic inverse seesaw mechanism \cite{Inverse} which has been
 also discussed by a number of authors
\cite{Inv:RNM:1986,Inv:Albright:1986,Inv:Witten:1986,Inv:Nandi:1986,Inv:Valle:1991,Inv:Wyler:LRS,Bernabeu:1987,Inv:Ma:1987} is essentially TeV scale seesaw. It has the
high potential to be directly verifiable at accelerator
energies and also by ongoing experiments on charged lepton flavor
violations \cite{lfvexpt1,lfvexpt2,lfvexpt:BaBar,lfvexpt3}.

 Even without taking recourse to string theories, in addition to the three
 RH neutrinos ($N_i, i=1,2,3$), one more  gauge singlet fermion per generation ($S_i,i=1,2,3$)
is added to the SM where the Lagrangian contains the $N-S$ mixing mass
term $M$. The heavy Majorana mass term is absent for RH neutrinos
which
turn out to be heavy pseudo-Dirac fermions. The introduction of
the global lepton symmetry breaking term in the Lagrangian, $\mu_SS^TS$,  
gives rise to the well known classic inverse seesaw formula \cite{Inverse}
\begin{equation}
m_{\nu}=\frac{M_D}{M}\mu_S(\frac{M_D}{M})^T, \label{invform}
\end{equation}
Naturally small value of $\mu_S$ in the $'$t Hooft sense \cite{tHooft:1975} 
brings down the inverse seesaw mechanism to the TeV scale
without having the need to fine tune the associated Dirac mass
matrix or Yukawa couplings. The presence of texture zeros in the
constituent matrices of different types of seesaw formulas have
been investigated consistent with neutrino oscillation data \cite{Mainak:2012,PRoy:2013,Mainak:2014,Ghosal:NP:2016,Ghosal:PLB:2016}.

 Recently models have been
  discussed using TeV scale heavy pseudo Dirac
neutrinos
\cite{mkp-sahoo:Proc:2014,
psb-rnm:2010,ap:2012,Das-Okada:2013,Kang:2016} where
dominant RH Majorana mass term $M_NNN$ is either absent in the
 Lagrangian or negligible. 
  In a contrasting situation \cite{Babu-Pati-Wil:2000,Lindner:schmidt-smirnov:2005}, when 
$\mu_S \sim M_{Planck}$, the seesaw scale $M\sim M_{GUT}$, and the model avoids the presence of any additional intermediate symmetry. While operating with the SM paradigm, it also dispenses with the 
larger Higgs representation ${126}_H\subset SO(10)$ in favour of much smaller one ${16}_H$
leading to the double-seesaw. The heavy RH neutrinos with $M_N\sim 10^{13}$ GeV, turn out to be Majorana fermions instead of being pseudo Dirac. 
While this is an attractive scenario in SUSY SO(10)\cite{Babu-Pati-Wil:2000}
, the coupling unification is challenging in the non-SUSY SO(10).
 
As discussed above if type-I seesaw is the neutrino mass mechanism at the TeV
scale, it must be appropriately suppressed either by finetuning or by
introducing textures to the relevant mass matrices\cite{RNM:rev16}. On the otherhand if type-II seesaw dominance in LR models or SO(10) is to account for neutrino masses, $W_R,Z_R$ boson masses must be at the GUT-PlancK scale in the prevailing dominance mechanisms 
\cite{Goh-rnm-nasri:2004,rnmmkp11}.\\  

 In view of this, it
would be quite interesting to explore, especially in the ccontext of
non-supersymmetric SO(10), possible new physics implications when
the would be dominant type-I seesaw cancels out exactly and analytically from the light
neutrino mass matrix even without needing any fine tuning or fermion
mass textures in $M_D$ and/or $M_N$ . The complete cancellation of type-I seesaw in the
presence of heavy RH Majorana mass term $M_NNN$ was explicitly proved
in ref.\cite{Kang-Kim:2006} in the context of SM extension when both
$N_i$ and $S_i$ are present manifesting in heavy RH neutrinos and
lighter singlet fermions. We call this as gauge singlet fermion assisted extended seesaw dominance mechanism. Since then the mechanism has been utilised in explaining baryon asymmetry of the universe via low-scale leptogenesis 
\cite{Kang-Kim:2006}, the phenomenon of dark matter (DM)
\cite{Cheon:2007} along  with cosmic ray anomalies
\cite{Cheon:2009}. More recently this extended seesaw mechanism for
neutrino masses in the SM extension has been exploited to explain the
keV singlet fermion DM along with low-scale leptogenesis
\cite{Kang-Patra:2014}.   

In the context of  LR intermediate scales in SUSY
SO(10), this mechanism has been applied to study coupling unification
and leptogenesis \cite{Majee:PLB:2007,mkp-arc:2010,Majee:PRD:2007} under gravitino constraint. Application to non-SUSY LR theory originating from
Pati-Salam model \cite{pp:2013} and non-SUSY SO(10) with TeV scale
$W_R,Z_R$ bosons have been made
\cite{app:2013,pas:2014} with the
predictions of a number of experimentally testable physical phenomena
by low energy experiments and including the observed dilepton excess
at LHC \cite{mkp-sahoo:NP:2015}. In these models the singlet fermion
assisted type-I seesaw cancellation mechanism operates and the extended seesaw (or inverse seesaw) formula dominates.\\ 

Following the standard lore in  type-II seesaw dominant models, the
dominant double beta decay rate in the $W_L-W_L$ channel is expected
to be dominated by the exchange of the LH Higgs triplet $\Delta_L$
carrying $|B-L|=2$. As such the predicted decay rate tends to be
negligible in the limit of larger Higgs triplet mass.  But it has been
shown quite recently  \cite{bpn-mkp:2015}, that the type-II seesaw dominance can occur assisted by
the gauge singlet fermion but with a phenomenal difference. Even for large LH
triplet mass in such models that controls the
type-II seesaw formula for light neutrino masses and mixings, the
  double
beta decay rate in the $W_L-W_L$ channel remains dominant as it is
controlled by the light singlet fermion exchanges. Other attractive
predictions are observable LFV decays, non-unitarity effects, and resonant leptogenesis mediated by TeV scale
quasi-degenerate  singlet fermions
 of the second and third generations. The model has been
noted to have  its
origin in non-SUSY SO(10) \cite{bpn-mkp:2015}. All the three types of gauge
singlet fermions in these models mentioned above are Majorana fermions on which we
focus in this review.

This article is organised in the following manner. 
In Sec.2 we explain how the Kang-Kim mechanism \cite{Kang-Kim:2006} operates within the SM paradigm
extended by singlet fermions.
In Sec.3 we show how a generalised neutral fermion mass matrix exists
in the appropriate extensions of the SM, LR theory, or SO(10).
 In  Sec.4 we 
 show emergence of the other dominant seesaw mechanism including the
 extended or inverse
 seesaw and type-II seesaw and cancellation of type-I seesaw. Predictions for LFV, CP violation and
non-unitarity effect is discussed in Sec.5. Predictions on double beta decay mediated
 by light singlet fermions in the
 $W_L-W_L$ channel is discussed in Sec.6 where we have given
 its mass limits from the existing experimental data. Applications to
 resonant leptogenesis mediated by TeV scale singlet fermions in MSSM and SUSY SO(10) are briefly discussed in Sec.7.
Singlet fermion assisted leptogenesis in non-SUSY SO(10) is discussed in Sec.8.
This work is summarised in Sec.9 with conclusions.

\section{MECHANISM OF EXTENDED SEESAW DOMINANCE IN SM EXTENSION}\label{sec.2}

Using the explicit derivation of Kang and Kim
\cite{Kang-Kim:2006}, here we discuss how the type-I contribution
completely cancels out paving the way for the dominance of extended seesaw
mechanism.
The SM is extended by introducing RH neutrinos $N_i(i=1,2,3)$ and an additional set of fermion singlets $S_i(i=1,2,3)$ , one for each generation.
After electroweak symmetry breaking, the Yukawa Lagrangian in the
charged lepton mass basis gives for the neutral fermions
\begin{eqnarray}
\mathcal{L}_{mass}= M_D \overline{\nu}N +\frac{1}{2}M_NN^{T}N
+M\overline{N}S  + h.c)
+\mu_S S^TS\label{smlag}
\end{eqnarray}
where $M_D=$ the Dirac neutrino mass matrix $=Y<\phi>$, $Y$ being the Yukawa matrix. This gives the $9\times 9$ neutral fermion mass matrix in the $(\nu, N^c, S)$ basis,
\ba
{\cal M}_\nu =  
\begin{pmatrix} 0 & M_D  & 0  \\
 M_D^T  & M_N & M^T\\ 0 & M &
 \mu_S \end{pmatrix}.  \label{fullnumatrix}
\ea
The type-I seesaw cancellation leading to dominance of extended seesaw
(or inverse seesaw) \cite{Inverse} proceeds in two steps: As $M_N>> M
>>M_D , \mu_S$, it is legitimate to integrate out the RH $N_i$ fields
at first
leading to the corresponding effective Lagrangian
\begin{eqnarray}
- \mathcal{L}_{\rm eff} &=& \left( M_D \frac{1}{M_N} M^T_D\right)_{\alpha \beta}\, \nu^T_\alpha \nu_\beta +
\left(M_{L} + M_D \frac{1}{M_N} M^T \right)_{\alpha m}\, \left(\overline{\nu_\alpha} S_m + \overline{S_m} \nu_\alpha \right)
\nonumber \\
&&\hspace*{4.0cm} +\left(M \frac{1}{M_N} M^T\right)_{m n}\, S^T_m S_n+\mu_SS^TS. 
\end{eqnarray}
Then diagonalisation of the $9\times 9$ neutral fermion mass matrix
including the result of  $\mathcal{L}_{\rm eff}$ gives 
conventional type-I seesaw term and another of opposite sign leading
to the cancellation. The light neutrino mass predicted is the same as
in the inverse seesaw case given in eq.(\ref{invform})\\ 

It must be emphasized that the earlier realizations of the classic inverse seesaw formula \cite{Inverse} were possible 
\cite{Inv:RNM:1986,Inv:Albright:1986,Inv:Witten:1986,Inv:Nandi:1986,Inv:Valle:1991,Inv:Wyler:LRS,Inv:Ma:1987}
with vanishing  RH Majorana mass $M_N=0$ in eq.(\ref{fullnumatrix}).  

Under the similar condition in which the type-I seesaw cancels out
the Majorana mass $m_S$ of the sterile neutrino and its mixing angle $\theta_S$ with light neutrinos are governed by 
\begin{eqnarray}
{ m}_S &=& \mu_S-M\frac{1}{M_N}M^T \sim -M\frac{1}{M_N}M^T, \nonumber\\
\tan 2\theta_S &=& 2\frac{M_D}{M}.\label{msform}
\end{eqnarray}
As $\mu_S$ is naturally small, it is clear that  type-I seesaw now
controls the gauge singlet fermion mass, although it has no role to play in
determining the LH neutrino mass.  These results have been shown to
emerge \cite{Majee-mkp:PLB:2007,mkp-sahoo:Proc:2014,mkp-sahoo:NP:2015,
bpn-mkp:2015,app:2013,pas:2014} from SO(10) with gauge fermion siglet extensions 
 by following the explicit
block diagonalisation procedure  in two steps while safeguarding the
hierarchy $M_N >> M > M_D, \mu_S$  with the supplementary condition $\mu_SM_N < M^2$.

\section{GENERALISED NEUTRAL FERMION MASS MATRIX} \label{sec.3}

A left-right symmetric (LRS) gauge theory $G_{2213D}(g_{2L}=g_{2R})$ at
higher scale ($\mu=M_P$)  is known to lead  to TeV scale asymmetric LR gauge theory 
$G_{2213}(g_{2L} \neq g_{2R})$ via D-Parity breaking
\cite{cmp1:1984,cmp2:1984,cmgmp:1985}. This symmetry further breaks to the SM gauge symmetry
by the VEV of the RH triplet $\Delta_R(1,3,-2,1)$ leading to massive
$W_R,Z_R$ bosons and RH neutrinos at the intermediate scale $M_R$. In
stead of $G_{2213D}(g_{2L}=g_{2R})$ it is possible to start  directly from  $SO(10)$  which has been discussed at length in a number of
investigations that normally leads to the type-I$\oplus$type-II hybrid seesaw
formula. In the absence of additional sterile neutrinos, the neutral
fermion matrix is standard $6\times 6$ form. Here we discuss how a generalised $9\times 9$ neutral fermion
mass matrix that  emerges in the presence of additional singlet fermions
contains the rudiments of various seesaw formulas.
As noted in Sec.1, the derivation of the minimal classic inverse seesaw mechanism \cite{Inverse} has been possible in theories gauge singlet fermion extensions of the SM are available\cite{Inv:RNM:1986,Inv:Albright:1986,Inv:Witten:1986,Inv:Nandi:1986,Inv:Valle:1991,Inv:Wyler:LRS,Bernabeu:1987,Inv:Ma:1987,Akhmedov:LRS-Linear-dyn:1995}.
Extensive applications of this mechanism have been discussed and reported in a
number of recent reviews
\cite{Valle:rev16,Morisi-Valle:2012,Deppisch:2012,de
  Gouvea:2013,Boucenna-Morisi-Valle:rev14,Deppisch:nuLHC:2015,Deppisch:nuColl:rev15}. Exploring possible
effects
on invisible Higgs decays
\cite{Joshipura-Valle:1993,Diaz-Valle:1998,Bonilla-Valle-Romao:2015},
prediction
of observable lepton flavor violation as a hall mark of the minimal
classic inverse seeaw mechanism has attracted considerable attention
earlier and during recent investigations
\cite{Heeck:2016,Deppisch-Valle:2005,Deppisch-Kosmas-Valle:2006,Deppisch-Desai-Valle:2014,Abada-Weiland:2014,Boucenna-Valle-Vicente:2015,Cirigliano:2004}. 
The effects of massive gauge singlet fermions have been found to be
consistent with electroweak precision observables \cite{EWprecision}
Earlier 
 its impact on a class of  left-right symmetric models have been examined
\cite{Akhmedov-Valle:1995,RNM-Kumar:1995,Akhmedov:LRSNJL:1996}. Prospects
of lepton flavor violation in the context of linear seesaw and
dynamical left-right symmetric model have been also investigated earlier\cite{Akhmedov:LRS-Linear-dyn:1995}.     

It is  well known that 15 fermions of one generation plus  a right
handed neutrino form the spinorial representation $16$ of SO(10) grand
unified theory \cite{georgi:1974}. In addition to three generation of
fermions ${16}_i (i=1,2,3)$, we also include one SO(10)-singlet fermion
per generation $S_i(i=1, 2, 3)$. We note that such singlets under the
LR gauge group or the SM can originate from the non-standard fermion
representations in SO(10) such as ${45}_F$ or ${210}_F$.

Under $G_{2213}$ symmetry the fermion and Higgs representations are, 

{\bf Fermions}\\\\
$Q_L$ $=$ $\begin{pmatrix}u\\d\end{pmatrix}_ L$ $\begin{pmatrix}2,1,1/3,3\end{pmatrix}$, 
$Q_R$ $=$ $\begin{pmatrix}u\\d\end{pmatrix}_ R$ $\begin{pmatrix}1,2,-1/3,3^{*}\end{pmatrix}$,\\
$L$ $=$ $\begin{pmatrix}\nu_l\\l\end{pmatrix}_L$ $\begin{pmatrix}2,1,-1,1\end{pmatrix}$,
$R$ $=$ $\begin{pmatrix}N_l\\l\end{pmatrix}_R$ $\begin{pmatrix}1,2,-1,1\end{pmatrix}$,\\
$S_i$ $=$$\begin{pmatrix}1,1,-1,1\end{pmatrix}$.\\

{\bf Higgs}\\\\
$\phi$ $=$ $\begin{pmatrix}\phi_1^{0}\hspace{0.5cm} \phi_{2}^{+}\\\phi_1^{-}\hspace{0.5cm}\phi_{2}^{0}\end{pmatrix}$$\begin{pmatrix}2,2,0,1\end{pmatrix}$,
$\Delta_L$ $=$ $\begin{pmatrix}\Delta_L^{+}/\sqrt{2} \hspace{0.5cm}\Delta_L^{++}\\\Delta_L^{0} \hspace{0.5 cm} -\Delta_L^{+}/\sqrt{2}\end{pmatrix}$
 $\begin{pmatrix}3,1,-2,1\end{pmatrix}$,\\\\
$\Delta_R$ $=$ $\begin{pmatrix}\Delta_R^{+}/\sqrt{2} \hspace{0.5cm}\Delta_R^{++}\\\Delta_R^{0} \hspace{0.5 cm} -\Delta_R^{+}/\sqrt{2}\end{pmatrix}$
 $\begin{pmatrix}1,3,-2,1\end{pmatrix}$,\\
$\chi_L$ $=$ $\begin{pmatrix}\chi_L^{0}\\\chi_L^{-}\end{pmatrix}$$\begin{pmatrix}2,1,-1,1\end{pmatrix}$,
$\chi_R$ $=$
        $\begin{pmatrix}\chi_R^{0}\\\chi_R^{-}\end{pmatrix}$$\begin{pmatrix}1,2,-1,1\end{pmatrix}$,\\
$\eta_o (1, 1, 0, 1)$,\\
where $\eta_o$ is a D-Parity odd singlet with transformation property
$\eta_o \to -\eta_o $ under $L\to R$. When this singlet acquires VEV
$<\eta_o>\sim M_P$, D-Parity breaks along with the underlying
left-right discrete symmetry but the asymmetric LR gauge theory
$G_{2213}$ is left unbroken down to the lower scales.  The  $G_{2213}$
gauge theory can further break down to the SM directly by the VEV of
RH Higgs triplet  $\Delta_R(1, 3, -2,1)\subset {126}_H\subset SO(10)$
or the RH Higgs doublet  $\chi_R(1, 2, -1,1)\subset {16}_H\subset SO(10)$.
 The D-Parity odd (even) singlets $\eta_o(\eta_e)$  were found to
 occur naturally in $SO(10)$ GUT theory \cite{cmp1:1984,cmp2:1984,cmgmp:1985}. Designating the quantum
 numbers of submultiplets under Pati-Salam symmetry $SU(2)_L\times
 SU(2)_R\times SU(4)_C$ ($\equiv G_{224}$), the submultiplet
 $(1,1,1)\subset {210}_H \subset SO(10)$ is $\eta_o$ where as  the submultiplet
 $(1,1,1)\subset {54}_H \subset SO(10)$ is $\eta_e$. Likewise the
 neutral component of the submultiplet $(1,1,15) \subset {45}_H\subset
 SO(10)$ behaves as $\eta_o$, but that in  $(1,1,15) \subset {210}_H$
 behaves as $\eta_e$. Thus the GUT scale symmetry breaking $SO(10) \to
G_{224D}$ can occur by the VEV of ${54}_H$ in the direction $<\eta_e>
\sim M_{GUT}$, but  $SO(10) \to
G_{224}$ can occur by the VEV of ${210}_H$ in the direction $<\eta_o>
\sim M_{GUT}$. Likewise  $SO(10) \to
G_{2213D}$ can occur by the VEV of the neutral component
$(1,1,0,1)_H\subset (1,1,15)_H \subset {210}_H$ , but  $SO(10) \to
G_{2213}$ can occur by the VEV of the neutral component of
$(1,1,0,1)_H\subset (1,1,15)_H \subset {45}_H$. As an example, one
minimal chain with TeV scale LR gauge theory  proposed recently in the
context of like-sign dilepton signals observed at LHC is,\\

\ba
SO(10) &
\stackrel {(M_U=M_P)}
{\longrightarrow}      
& SU(2)_L \times
SU(2)_R \times U(1)_{B-L}\times  SU(3)_C~~[G_{2213}] \nonumber \\ 
&
\stackrel {(M_R)}
{\longrightarrow}& SU(2)_L \times
U(1)_Y\times  SU(3)_C~~[\rm SM] \nonumber\\
&
\stackrel {(M_Z)}{\longrightarrow}&SU(3)_C \times U(1)_Q . \label{minchain}
\ea

In this symmetry breaking pattern all LH triplets and doublets are
near the GUT scale, but RH triplets or doublets are near the
$G_{2213}$ breaking intermediate scale $M_R$ which could be $\sim ({\rm
  few}- 100)$ TeV. Out of two minimal  models with GUT scale D-Parity
breaking satisfying the desired decoupling criteria  $M_N >>  M >> M_D, \mu_S$ r \cite{mkp-sahoo:NP:2015,bpn-mkp:2015}, dominance
of extended seesaw  in the presence of gauge singlet fermions
 has been possible in ref.\cite{mkp-sahoo:NP:2015} with single $G_{2213}$
intermediate scale corresponding to TeV scale $W_R,Z_R$ bosons. The
extended seesaw dominance in the presence of fermion singlets in
SO(10)  have been also realised 
 including additional intermediate symmetries $G_{2214D}$ and
 $G_{224}$ where observable proton decay, TeV scale $Z_R$ boson and RH
 Majorana neutrinos, observable proton decay,
  $n- {\bar n}$ oscillation, and rare kaon decay have been predicted. 
Interestingly the masses of $W_R$ boson and lepto-quark gauge bosons
of $SU(4)_C$ 
 have been predicted at $\sim 100$ TeV which could be accessible to planned LHC
at those energies  where $W_R$ boson scale $\sim (100-1000)$ TeV
matching with observable $n- {\bar n}$ oscillation and rare kaon decay
has been predicted. But the heavy RH neutrino and $Z_R$ boson scales
being near TeV scale
have been predicted to be accessible to LHC and planned accelerators
\cite{app:2013,pas:2014}. That non-SUSY GUTs with two-intermediate
scales permit a low mass $Z_R$ boson was noted much earlier \cite{mkp-ARC:1982}.

 In eq.(\ref{minchain}), instead of breaking directly to SM, the 
$G_{2213}$ breaking may occur in two steps  $G_{2213}\to G_{2113} SM$
where  $G_{2113}$ represents the gauge symmetry $SU(2)_L \times
U(1)_R \times U(1)_{B-L}\times  SU(3)_C~~[G_{2113}]$. This
promises the interesting possibility of TeV scale $Z_R$ boson with the
constraint $M_{W_R} >> M_{Z_R}$. Thus the model can be discriminated
from the direct LR models if $Z_R$ boson is detected at lower mass scales
than the $W_R$-boson. There are currently ongoing
accelerator searches for this extra heavy neutral gauge boson. This has
been implemented recently with type-II seesaw dominance in the
presence of added fermion singlets \cite{bpn-mkp:2015}. As we will discuss below both these
types of models predict light neutrinos capable of mediating double beta    
 decay rates in the $W_L-W_L$ channel saturating the current
 experimental limits. In addition resonant leptogenesis mediated by
 heavy sterile neutrinos has been realised in the model of
 ref.\cite{bpn-mkp:2015}.

The $G_{2213}$ symmetric  Yukawa Lagrangian descending from $SO(10)$ symmetry can be written as
\begin{eqnarray}
\mathcal{L}_{\rm Yuk} &= & \sum_{i=1,2}Y_{i}^{\ell} \overline{\psi}_L\, \psi_R\, \Phi_{i} 
                       + f\, (\psi^c_R\, \psi_R \Delta_R +\psi^c_L\, \psi_L \Delta_L)
                       + y_{\chi}\, (\overline{\psi}_R\, S\, \chi_R +\overline{\psi}_L\, S\, \chi_L)\nonumber \\
                       &+\text{h.c.},\label{Yuk-Lag} 
\end{eqnarray}
where $\Phi_{1,2}\subset {10_{{H_1},{H_2}}}$ are two bidoublets, $(\Delta_L,\Delta_R)\subset {126}_F$ and $(\chi_L,\chi_R)\subset
{16}_H$.

 Including the induced VEV contribution to $\Delta_L$, the
Yukawa mass
term can be written as 
\begin{eqnarray}
\mathcal{L}_{mass}= (M_D \overline{\nu}N +\frac{1}{2}M_NN^{T}N
+M\overline{N}S +M_L\overline{\nu_L}S + h.c)
+m_{\nu}^{II}{\nu^T}\nu+\mu_S S^TS.
\end{eqnarray}
Here the last term denotes the gauge invariant singlet mass term where
naturalness criteria demands $\mu_S$ to be a very small parameter.
In the $(\nu, S, N^C)$ basis   
the  generalised form of the $9\times 9$ neutral fermion mass matrix after electroweak symmetry breaking can be written as 

\ba
{\cal M}_\nu =  
\begin{pmatrix} m_{\nu}^{II} & M_L & M_D  \\
M_L^T & \mu_S & M^T\\ M_D^T & M & M_N \end{pmatrix},  \label{fullnumatrix}
\ea
where $M_D=Y\langle \Phi\rangle$, $M_N=fv_R$, ${\rm
  M}=y_{\chi}\langle\chi_R^0\rangle$,$M_L=y_{\chi}\langle\chi_L^0\rangle$.
In this model the symmetry braking mechanism and the VEVs are such
that $M_N > M \gg M_D$. The LH triplet scalar mass $M_{\Delta_L}$ and
RH neutrino masses being at the the heaviest mass scales in the
Lagrangian, this triplet scalar field and the RH neutrinos are at first integrated out
leading to the effective Lagrangian at lower scales \cite{Kang-Kim:2006,bpn-mkp:2015},
\begin{eqnarray}
- \mathcal{L}_{\rm eff} &=& \left(m_{\nu}^{II} + M_D \frac{1}{M_N} M^T_D\right)_{\alpha \beta}\, \nu^T_\alpha \nu_\beta +
\left(M_{L} + M_D \frac{1}{M_N} M^T \right)_{\alpha m}\, \left(\overline{\nu_\alpha} S_m + \overline{S_m} \nu_\alpha \right)
\nonumber \\
&&\hspace*{4.0cm} +\left(M \frac{1}{M_N} M^T\right)_{m n}\, S^T_m S_n+\mu_SS^TS. 
\end{eqnarray}

\section{CANCELLATION OF TYPE-I SEESAW AND DOMINANCE OF OTHERS} \label{sec.4}
\par\noindent{\bf(a) Cancellation of Type-I Seesaw}\\
Whereas the heaviest RH neutrino mass matrix $M_N$ separates out trivially, the other two
  $3\times 3$ mass matrices $\mathcal{M}_\nu$, and  $\mathcal{M}_S$
are extracted through various steps of block diagonalisation. The
details of various steps are given in refs. \cite{bpn-mkp:2015,
app:2013,pas:2014} 
\begin{eqnarray}
 \mathcal{M_\nu} &=&  m_{\nu}^{II} + \left(M_D M_N^{-1} M^T_D\right) -
(M_D M_N^{-1} M^T_D )+ M_L(M^T M_N^{-1}M)^{-1} M_L^{T}\nonumber \\
 &&-M_L(M^T M_N^{-1}M)^{-1}(M^T M_N^{-1}M_D^T) -(M_D M_N^{-1} M) (M^T
 M_N^{-1}M)^{-1} M_L^{T}\nonumber\\
&&+M_DM^{-1}\mu_S{M_DM^{-1}}^T, \nonumber\\
 \mathcal{M_{\mathcal {S}}} &=&\mu_S-MM_N^{-1}M^T+.... ,\nonumber\\
 \mathcal{M_{\mathcal {N}}} &=& M_N.\label{massmatrices} 
 \end{eqnarray} 
From the first of the above three equations, it is clear that the
type-I seesaw term cancels out  with another of opposite sign
resulting from block diagonalisation. Then the generalised form of the  light neutrino mass matrix turns out to be
 \ba
{\cal M}_{\nu} &=&~ fv_L+M_LM^{-1}M_N(M^T)^{-1}M_L^T \nonumber\\
&&-[M_LM^{-1}M_D^T
  +M_D(M_LM^{-1})^T]+\frac{M_D}{M}\mu_S(\frac{M_D}{M})^T.
 \label{hybrid}
\ea
In different limiting cases this generalised light neutrino mass
matrix reduces to the corresponding well known neutrino mass
formulas.\\
\par\noindent{\bf(b). Linear Seesaw and Double Seesaw}\\
 With $M_L=y_{\chi}v_{\chi_L}$ that induces $\nu-S$ mixing, the second
 term in  eq.(\ref{hybrid}) is the double seesaw formula,
\be
{\cal M}^{(\rm double)}_{\nu}=M_LM^{-1}M_N(M^T)^{-1}M_L^T
.\label{double}
\ee
The third term in eq.(\ref{hybrid}) represents the linear seesaw
formula
\be
{\cal M}^{(\rm linear)}_{\nu}=-[M_LM^{-1} M_D^T+M_D(M_LM^{-1})^T].\label{linear}
\ee
Similar formulas have been shown to emerge from single-step breaking
of SUSY GUT models \cite{Linear:Barr:2004,Linear:Fukuyama:2007} which require
the presence of three gauge singlet fermions.

Using the D-Parity breaking mechanism of ref.\cite{cmp1:1984,cmp2:1984},
an interesting model of linear see saw mechanism in the context of
supersymmetric SO(10) with suceessful gauge coupling unification 
\cite{Linear:Malinsky:2005} has been suggested in the presence of three gauge
singlet fermions. A special feature of this linear seesaw, compared to
others \cite{Linear:Barr:2004,Linear:Fukuyama:2007} is that the neutrino mass
formula is suppressed by the
SUSY GUT scale but it is decoupled from the low $U(1)_{B-L}$ breaking
scale. In addition to prediction of TeV scale superpartners, the model
provides another important  testing ground through manifestation of extra $Z'$
boson at LHC or via low-energy neutrino scattering experiment
\cite{Garces-Valle:2012}.

\par\noindent{\bf (c). Type-II Seesaw}\\
When the assigned or induced VEV $<\chi_L>=0$, or negligible and
$\mu_S \to 0$, in eq.(\ref{invform}), type-II seesaw dominates leading to
\begin{equation}
m_{\nu} \simeq  fv_L .\label{typeIIform}   
\end{equation}
As noted briefly in Sec.1, in the conventional models \cite{Goh-rnm-nasri:2004,rnmmkp11}  of
type-II seesaw dominance in SO(10), the $W_R,Z_R$ boson masses have to be at the
GUT-Planck scale. As a phenomenal development, this singlet-fermion 
assisted type-II seesaw dominance permits $U(1)_{B-L}$ breaking
scale associated with $G_{2213}$ or $G_{2113}$ breaking (i,e the $W_R,Z_R$ boson masses) accessible to accelerator
energies including LHC. At the same time
the heavy  $N-S$ mixing mass terms $M_i (i=1,2,3)$ at the TeV scale are capable of
mediating observable LFV decay rates closer to their current
experimental values \cite{lfvexpt1,lfvexpt2,lfvexpt:BaBar,lfvexpt3} as discussed in Sec.\ref{sec.5}.
 Consequences of this new Type-II seesaw dominance with  TeV scale
$Z_R$ boson mass has
been investigated in detail \cite{bpn-mkp:2015} in which charged
triplet mediated LFV decay rates are negligible but singlet fermion
decay rates are observable. Also predictions of observable double beta decay rates close
to their experimental limits are discussed below in Sec.\ref{sec.6}. While the principle
of such a dominance is clearly elucidated in this derivation, the
details of the model with TeV scale $G_{2213}$
symmetry will be reported elsewhere.

\par\noindent{\bf (d). Extended Seesaw}\\
It is quite clear that the classic inverse see saw formula \cite{Inverse} of eq.(\ref{invform}) for light neutrino
mass emerges when the  LH triplet mass is large and the VEV $<\chi_L>=0$
which is possible in a large class of non-SUSY  models with left-right,
Pati-salam, and SO(10)  gauge groups with $D-Parity$ broken at high
scales \cite{cmp1:1984,cmp2:1984,cmgmp:1985} with $M_N=0$ leading to
RH neutrinos as heavy pseudo Dirac fermions. Particularly in SO(10)
some non-SUSY examples are \cite{ap:2012,mkp-sahoo:Proc:2014} and SUSY examples are \cite{psb-rnm:2010,Blanchet:2010}  non-SUSY SO(10) exa even without taking
recourse to superstring theories provided we reconcile with gauge
hierarchy problem and fine-tuned non-SUSY models \cite{mkp-sahoo:Proc:2014,mkp-sahoo:NP:2015,
app:2013,pp:2013}.\\

 As noted in Sec.1, the derivation of classic inverse seesaw
mechanism \cite{Inverse,Inv:RNM:1986,Inv:Albright:1986,Inv:Witten:1986,Inv:Nandi:1986,Inv:Valle:1991,Inv:Wyler:LRS,Bernabeu:1987,Inv:Ma:1987}
 has $M_N=0$ in eq.(\ref{fullnumatrix}). More recent applications in
 LRS and GUTs have been  discussed with relevant reference to earlier
 works in \cite{Lee-psb-rnm:2013,ap:2012,Das-Okada:2013,
Vicente:2010,Abada-Lucente:2014,Weiland:2014,Lindner:2016,Bonilla:2016}.

In this section  we have discussed that, in spite of the presence of the heavy Majorana mass term of RH neutrino,  each of the three seesaw
mechanisms : (i) Extended Seesaw, (ii) Type-II seesaw ,  (iii)
Linear Seesaw or Double seesaw, can dominate as light neutrino mass ansatz when the
respective limiting conditions are  satisfied. Also the seesaw can operate
in the presence of TeV scale $G_{2213}$ or $G_{2113}$ gauge symmetry originating
from non-SUSY SO(10) \cite{bpn-mkp:2015,app:2013,pas:2014}. As the TeV
scale theory spontaneously breaks to low-energy theory $U(1)_{em}\times
SU(3)_C$ through the electroweak symmetry breaking of the standard
model, these seesaw mechanisms are valid in the SM extensions with
suitable Higgs scalars and three generations of $N_i$ and $S_i$. For
example without taking recourse to LR gauge theory type-II seesaw
can be embedded into the SM extension by inclusion of LH Higgs triplet  
$\Delta_L(3,-2,1)$ with $Y=-2$. The induced VEV can be generated by
the trilinear term $\lambda
M_{tr}.\Delta_L^{\dagger}\phi^{\dagger}\phi^{\dagger}$ \cite{Ma-Sarkar}. The origin of
such induced VEV in the direct breaking of $SO(10) \to SM$ is well
known.\\

 \par\noindent{\bf (e). Hybrid Seesaw }\\
In the minimal SO(10), without extra fermion singlets, one example of
hybrid seesaw with type-I$\oplus$type-II is given in eq.(\ref{comb}). 
There are a number of investigations where this hybrid seesaw has been
successful in parametrising small neutrino masses with large mixing
angles along with  $\theta_{13} \sim 8^o$ in SUSY SO(10)
\cite{Goh-rnm-ng:2003,typeIISO10}  and LR models.
But the present mechanism of type-I seesaw cancellation suggests a possible new  hybrid
seesaw formula as a combination of type-II$\oplus$Linear$\oplus$Extended
seesaw as revealed from eq.(\ref{hybrid}). Neutrino physics
phenomenology may yield interesting new results with this new
combination with additional degrees of freedom to deal with neutrino
oscillation data and leptogenesis covering coupling unification in
SO(10) which has a very rich structure for dark matter.

Using the D-Parity breaking mechanism of ref.\cite{cmp1:1984,cmp2:1984},
an interesting model of linear see saw mechanism in the context of
supersymmetric SO(10) with suceessful gauge coupling unification 
\cite{Linear:Malinsky:2005} has been suggested in the presence of three gauge
singlet fermions. A special feature of this linear seesaw, compared to
others \cite{Linear:Barr:2004} is that the neutrino mass formula is suppressed by the
SUSY GUT scale but seedecoupled from the low $U(1)_{B-L}$ breaking scale
which can be even at $\sim {\rm few}$ TeV. 
This serves as a testing ground through manifestation of extra $Z'$
boson at LHC or via low-energy neutrino scattering experiments \cite{Garces-Valle:2012}. Being a SUSY model it also predicts TeV
scale superpartners expected to be visible at LHC. 

\par\noindent{\bf (f). Common Mass Formula for Sterile Neutrinos}\\ 
In spite of different types of seesaw formulas in the corresponding
limiting cases the formula for sterile neutrino mass remains the same
as in eq.(\ref{msform}) which does not emerge from the classic inverse seesaw approach with $M_N=0$

We conclude this section by noting that the classic inverse seesaw
mechanism was gauged at the TeV scale through its embedding in
non-SUSY SO(10) with the prediction of experimentally accessible Z'
boson, LFV decays, and non-unitarity effects \cite{ap:2012}.
The possibility of gauged and extended inverse seesaw mechanism with dominant contributions to both lepton flavor and lepton number non-conservation  was at first  noted in the context Pati-Salam model
in ref.\cite{pp:2013} and in the context of non-SUSY SO(10) in
ref.\cite{app:2013,pas:2014} with type-I seesaw cancellation. The generalised form of hybrid seesaw of
eq.(\ref{hybrid}) in non-SUSY SO(10) with type-I cancellation was realised in ref.\cite{bpn-mkp:2015}. As a special case of this model, the experimentally verifiable phenomena like extra $Z'$ boson, resonant leptogenesis, LFV decays, and double beta decay rates closer to the current 
search limits  were decoupled from the intermediate scale type-II seesaw dominated neutrino mass 
genaration mechanism. Proton lifetime prediction for $p\to e+\pi^0$ mode also turns out to be within the accessible range.

\section{PREDICTIONS FOR LFV DECAYS, CP VIOLATION, AND NONUNITARITY}\label{sec.5}
The presence of non-vanishing  neutrino masses with generational mixing evidenced from the oscillation data \cite{nudata}, in principle, induce charged lepton flavor violating (LFV) decays. For a recent review see   \cite{Morisi-Valle:2012} and references therein.The observed neutrino mixings through weak charged currents lead to non-conservationn of lepton flavor numbers $L_e$, $L_{\mu}$, and $L_{\tau}$ resulting in the predictions of $\mu \to e + \gamma$, $\tau \to e + \gamma$, $\tau \to \mu + \gamma$, $\mu \to ee{\bar e}$, and a host of others    
 \cite{lfvexpt1,lfvexpt2,lfvexpt:BaBar,lfvexpt3}. If the non-SUSY SM is minimally
 extended to embrace tiny netrino masses and mixings through GIM
 mechanism as the only underlying source of charged lepton flavor
 violation, the loop mediated branching ratio is
\ba
\text{Br}\left(\ell_\alpha \rightarrow \ell_\beta + \gamma \right) \simeq
 \frac{\alpha}{32\pi}|\sum_{j=2,3} {U_{\alpha i}}{U_{\beta j}}^*\frac{{\Delta m}_{j1}^2}{M^2_W}|^2.\label{SMBR}
\ea
These branching ratios turn out to be $\le 10^{-53}$ ruling out any possibility
of experimental observation of the decay rates.
In high intermediate scale non-SUSY SO(10) models with Dirac
neutrino mass matrix $M_D$ similar to the up-quark mass matrix $M_u$,
the  type-I seesaw ansatz for neutrino masses constrains the heavy RH
neutrino masses $M_N \ge 10^{12}$ GeV resulting in branching ratio values $\le 10^{-40}$ which are
far below the current experimental limits. Another drawback of the
model is that the underlying neutrino mass generation mechanism and the
predicted $W_R$ boson mass can not be verified directly.

On the other hand SUSY GUTs  are well
known to provide  profound predictions of CP-violations and LFV decay branching ratios
closer to the current experimental limits  inspite of their high scale
seesaw mechanisms for neutrino masses. Some of the extensively available reviews on this subject are 
\cite{Babu:NOON2004,Morisi-Valle:2012,Deppisch:2012,Lindner:2016,Deppisch-PSB-Pilaftsis:rev15}. The superpartner
masses near $100-1000$ GeV are necessary for such predictions. 

As profound
applications of the classic inverse seesaw mechanism it has been noted
that the presence of heavy pseudo Dirac fermions would
manifest through LFV decays
\cite{Bernabeu:1987,Deppisch-Valle:2005,Deppisch-Kosmas-Valle:2006,Deppisch-Desai-Valle:2014,Abada-Weiland:2014,Boucenna-Valle-Vicente:2015}
and also in lepton number violation \cite{Valle-Singer:1983,Valle:1983}. They
are also likely to contribute to the modifications of the electroweak
observables \cite{EWprecision} keeping them within their allowed limits.  It has
been also emphasized  that the SUSY inverse seesaw mechanism for
neutrino masses further enhances the  LFV decay
rates \cite{Deppisch-Valle:2005}.
As a  direct test of the seesaw mechanism, these heavy particles with
masses 
near the TeV scale can be produced at high energy colliders including
LHC
\cite{mkp-sahoo:Proc:2014,mkp-sahoo:NP:2015,bpn-mkp:2015,Das-Okada:2013,psb-rnm:2010}. Other
interesting signatures have been reviewed in ref.\cite{Valle:rev16}.  

More interestingly a  linear seesaw formula has
been predicted from supersymmetric SO(10) with an extra $Z'$ boson
mass accessible to LHC \cite{Linear:Malinsky-Valle:2005}. The TeV
scale classic inverse seesaw mechanism and $W_R$ gauge boson masses  have been embedded in SUSY SO(10)
with rich struture for leptonic CP violation, non-unitarity effects, and LFV
decay branching ratios accessible to ongoing experiments \cite{psb-rnm:2010}. The impact of such a model with TeV scale pseudo Dirac type RH neutrinos has been investigated on proton lifetime predictions and leptogenesis \cite{Blanchet:2010} 

As supersymmetry has not been experimentally observed so far, an
interesting conceptual and practical issue is to confront LFV decay
rates accessible to ongoing experimental searches along with the
observed tiny values of light neutrino masses. In this section we
summarize how, in the absence of SUSY,    
the classic inverse seesaw and the extended seesaw could still serve
as powerful mechanisms to confront  neutrino mass, observable lepton
flavor violation \cite{mkp-sahoo:Proc:2014,ap:2012} and, in addition, dominant lepton
number violation \cite{mkp-sahoo:NP:2015,pp:2013,app:2013,pas:2014} in non-SUSY
SO(10).  The possibilities of detecting TeV scale $W_R$ bosons have
been also explored recently  in non-SUSY SO(10)
\cite{psb-rnm:so10:2015} and in
\cite{mkp-sahoo:NP:2015,
pp:2013,app:2013,pas:2014}. 
Besides earler works in  left-right-symmetric model 
\cite{Sarkar:LRS} and those cited in Sec.1-Sec.4, more 
recent works include
\cite{Kadastic:dm,
Gluza:2016,Deppisch:LRS-1:2016,Deppisch:LRS-2:2015,
Rodejohann:2013,Dobrescu:2015,Das-Valle:2012}, and in references cited in these papers.\\

In contrast to negligible LFV decay rates and branching ratios predicted in the
non-SUSY SM modified by GIM mechanism, we discuss in Sec.5.2 how the
non-SUSY SO(10) predicts the branching ratios in the range
$10^{-13}-10^{-16}$ consistent with small neutrino masses dictated by
classic inverse seesaw,
extended inverse seesaw or Type-II seesaw in the presence of added
fermion singlets.

\subsection{Neutrino Mixing and Non-Unitarity Matrix} 

The light neutrino flavor state is now a mixture of  three mass eigen
states ${\hat \nu}_i, {\hat s}_i$ and ${\hat N}_i$
\begin{equation}
\nu_{\alpha}= U_{\alpha,i}{\hat \nu}_i+{\cal V}_{\alpha,i}{\hat
  s}_i+{\cal N}_{\alpha,i} {\hat N}_i, \label{supstate}
\end{equation}
where in the diagonal bases of $S_i$ and $N_i$  ${\cal V}_{\alpha,i}=
(M_D/M)_{\alpha,i}$ and  ${\cal N}_{\alpha,i}=
(M_D/M_N)_{\alpha,i}$. In cases where the matrices $M, M_N$ are non-diagonal,
the corresponding flavor mixing matrices are taken as additional factors to
define $\nu-S$ or $\nu-N$ mixing matrices.  
 For the sake of simplicity,
 treating the $N-S$ mixing mass matrix $M$ as diagonal
\ba
M ={\rm diag}(M_1,M_2,M_3),\label{Mdiag}
\ea
and, under the assumed hierarchy $M_N >> M$, the formula for the
non-unitarity deviation matrix element $\eta_{\alpha \beta}$ has been defined
 in the respective cases \cite{mkp-sahoo:NP:2015,bpn-mkp:2015,
pp:2013,app:2013,pas:2014,Ilakovac,Malinsky}
\ba
\eta &=& \frac{1}{2}X.X^{\dag}=M_DM_R^{-2}M_D^{\dag},\nonumber\\
\eta_{\alpha\beta}&=&\frac{1}{2}\sum_{k=1,2,3}\frac{M_{D_{\alpha k}}M_{D_{\beta
    k}}^*}{M_{k}^2}.\label{etaele}
\ea
The Dirac neutrino mass matrix $M_D$ needed for the fits to neutrino
oscillation data through extended seesaw formula and prediction of
nonunitarity effects has been derived from 
the GUT scale fit of charged fermion masses in the case of non-SUSY
SO(10)\cite{mkp-sahoo:NP:2015,bpn-mkp:2015,
ap:2012,app:2013,pas:2014}. For
this purpose, the available data at the electroweak scale on charged
fermion masses and mixings are extrapolated to the GUT scale
\cite{Das-Parida:2001}. The fitting is done following the method of
ref.\cite{Babu-rnm:1993} by suitably adding additional contributions due to
VEVs of additional bi-doublets or higher dimensional operators,
wherever necessary. In the inverse seesaw case with almost degenerate heavy pseudo
Dirac neutrinos, $M_D$ has been derived in the case of SUSY SO(10) with
TeV scale $G_{2213}$ symmetry \cite{psb-rnm:2010} and in non-SUSY
SO(10) with TeV scale $G_{2113}$ symmetry\cite{ap:2012}. In the case
of extended seesaw dominance in non-SUSY SO(10) it has been derived in
ref.\cite{mkp-sahoo:NP:2015,
pp:2013,app:2013,pas:2014} whereas for type-II seesaw
dominance they it has been derived in ref.\cite{bpn-mkp:2015}.       
The value of $M_D$ thus derived at the GUT scale is extrapolated to
the TeV scale following the top-down approach. It turns out that 
such values are approximately equal to the one shown in the following section at 
eq.(\ref{MDatMR0}).

For the general non-degenerate case of the $M$ matrix, ignoring the
heavier RH neutrino contributions, and saturating the
upper bound  $|\eta_{\tau\tau}|<2.7\times 10^{-3}$ gives
\ba
\eta_{\tau \tau}&=&\frac{1}{2}\left[\frac{0.1026}{M_1^2}+\frac{7.0756}{M_2^2}+\frac{6762.4}{M_3^2}\right]\nonumber\\
 &=& 2.7\times 10^{-3}.\label{costr}
 \ea
By inspection this equation gives the lower bounds\\

$M_1 > 4.35~(\rm GeV),~M_2 > 36.2(\rm GeV), M_3 > 1120 ~(\rm GeV)$.\label{M123bounds}  
and for the degenerate case $M_{Deg}=1213$ GeV.  
$M (1213, 1213, 1212)$ GeV. For the partial degenerate case of $M_1=M_2\neq M_3$ 
the solutions can be similarly derived as in ref\cite{bpn-mkp:2015,
app:2013} and one example is $M (100,100,1319.67)$~GeV. .

Experimentally constrained  lower bounds of the non-unitarity matrix elements are
\ba
|\eta_{\tau\tau}|& \le & 2.7\times 10^{-3},~~|\eta_{\mu\mu}|\le 8.0\times 10^{-4},\nonumber\\
|\eta_{ee}|& \le &2.0\times 10^{-3},~~|\eta_{e\mu}|\le 3.5\times
 10^{-5},\nonumber\\
|\eta_{e\tau}| & \le & 8.0\times 10^{-3},~~|\eta_{\mu\tau}| \le  5.1\times
10^{-3}.\label{etabnd}
\ea
Out of several estimations of the elements of $\eta$-matrx
\cite{mkp-sahoo:NP:2015,bpn-mkp:2015,ap:2012,pp:2013,app:2013,pas:2014}
carried out in non-SUSY SO(10), here we give one example of ref.\cite{bpn-mkp:2015}. 
Using the Dirac neutrino mass matrix from ref.\cite{bpn-mkp:2015} and
allowed solutions of $M_i$, the values of the $\eta_{\alpha\beta}$ parameters and
their phases as functions of $M_i$ are determined using
eq.(\ref{etaele}). These results are 
presented in Table  \ref{tabeta}. 
\begin{table*}
\begin{center}
\begin{tabular}{|c|c|c|c|c|c|c|c|c|}
\hline
$m_{R_1}=m_{R_2}$&$m_{R_3}$&$|\eta_{e\mu}|$&$\delta_{e\mu}$&$|\eta_{e\tau}|$&$
\delta_{e\tau}$&$|\eta_{\mu\tau}|$&$\delta_{\mu\tau}$\\
(GeV)&(GeV)&&&&&&\\ \hline
$1213.11$&$1213.11$&$2.737\times 10^{-8}$&$1.920$&$4.543\times 10^{-7}$&$1.78$&$2.318\times
10^{-5}$&$2.391\times 10^{-7}$ \\ \hline
$500$&$1280$&$8.132\times 10^{-7}$&$1.326$&$3.746\times 10^{-7}$&$1.456$&
$2.426\times 10^{-5}$&$2.723\times 10^{-5}$ \\ \hline
$100$&$1119.67$&$6.543\times 10^{-6}$&$0.728$&$4.834\times 10^{-6}$&$0.974$&
$2.975\times 10^{-5}$&$8.932\times 10^{-4}$ \\ \hline
$50$&$1545.31$&$7.652\times 10^{-6}$&$0.203$&$9.754\times 10^{-6}$&$0.342$&
$3.424\times 10^{-5}$&$2.813\times 10^{-3}$\\ \hline
\end{tabular}
\end{center}
\caption{Predictions of moduli and phases of nonunitarity matrix $\eta_{\alpha\beta}$  as a
   function of allowed values of masses $M_1,M_2$, and $M_3$. The
   Dirac-neutrino matrix is the same as in ref.\cite{bpn-mkp:2015}}
\label{tabeta}
\end{table*}

\subsection{LFV Decay Branching Ratios vs. Neutrino Mass}
The most important outcome of non-unitarity effect is expected to manifest 
through ongoing experimental searches for LFV decays such as  $\tau\to
e\gamma$, $\tau\to \mu\gamma$ ,$\mu\to e\gamma$
\cite{lfvexpt1,lfvexpt2,lfvexpt:BaBar,lfvexpt3,Heeck:2016}. In
left-right symmetric models the LFV decay contribution due to
$W_R-N_i(i=1,2)$ mediaton has been computed as early as 1981 assuming
a GIM like mechanism in the RH sector \cite{Riazuddin:1981}.
 Some of the current
experimental bounds on the branching ratios are Br$(\mu^+ \to e^+
\gamma) < 5.7 \times 10^{-13}$,  Br$(\mu^+ \to e^+
\gamma) < 5.7 \times 10^{-13}$, Br$(\mu^+ \to e^+
\gamma) < 5.7 \times 10^{-13}$,  Br$(\mu^+ \to e^+
\gamma) < 5.7 \times 10^{-13}$. An uptodate list of experimental
results on various LFV processes and their future projections have
been summarised in \cite{Heeck:2016}. In these models
contribution to the branching ratios due to the heavier the RH neutrinos is
subdominant compared to the lighter singlet fermions 
\begin{eqnarray}
\label{eq:LFV}
\text{Br}\left(\ell_\alpha \rightarrow \ell_\beta + \gamma \right) =
          \frac{\alpha^3_{\rm w}\, s^2_{\rm w}\, m^5_{\ell_\alpha}}
          {256\,\pi^2\, M^4_{W}\, \Gamma_\alpha} 
           \left|\mathcal{G}^{N}_{\alpha \beta} + \mathcal{G}^{S}_{\alpha \beta}\right|^2 \, ,
\end{eqnarray}
\begin{eqnarray}
&\text{where}~&\mathcal{G}^{N}_{\alpha \beta} =
        \sum_{k} \left(\mathcal{V}^{\nu\, N}\right)_{\alpha\, k}\, 
         \left(\mathcal{V}^{\nu\, N}\right)^*_{\beta\, k} 
         \mathcal{I}\left(\frac{m^2_{N_k}}{M^2_{W_L}}\right) \, ,
         \nonumber \\
& & \mathcal{G}^{S}_{\alpha \beta} = \sum_{j} \left(\mathcal{V}^{\nu\, S}\right)_{\alpha\, j}\, 
         \left(\mathcal{V}^{\nu\, S}\right)^*_{\beta\, j} 
         \mathcal{I}\left(\frac{m^2_{S_j}}{M^2_{W_L}}\right) \, , \nonumber \\
&{\rm and} & \mathcal{I}(x) = -\frac{2 x^3+ 5 x^2-x}{4 (1-x)^3} 
                - \frac{3 x^3 \text{ln}x}{2 (1-x)^4}\, .\nonumber \\
\end{eqnarray}
 Because of the condition $M_N>> M$, the RH neutrino exchange contribution is however damped out compared to the sterile fermion singlet contributions. Using 
allowed solutions for $M_1,M_2,M_3$, our estimations in the partial degenerate case are given in Table \ref{tabcpv}.

\begin{figure}[htb!]
\includegraphics[width=6cm,height=6cm]{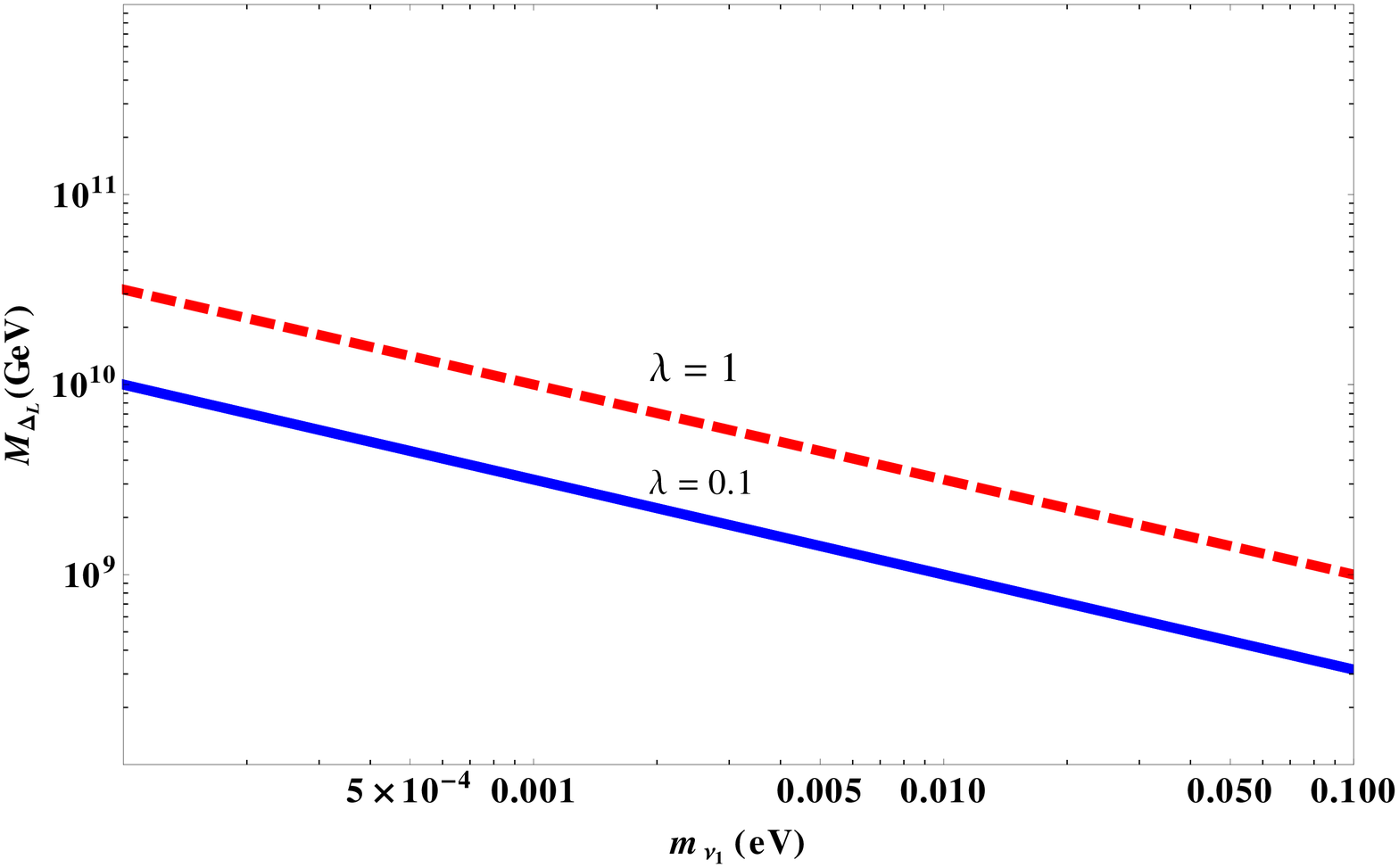}
\includegraphics[width=6cm,height=6cm]{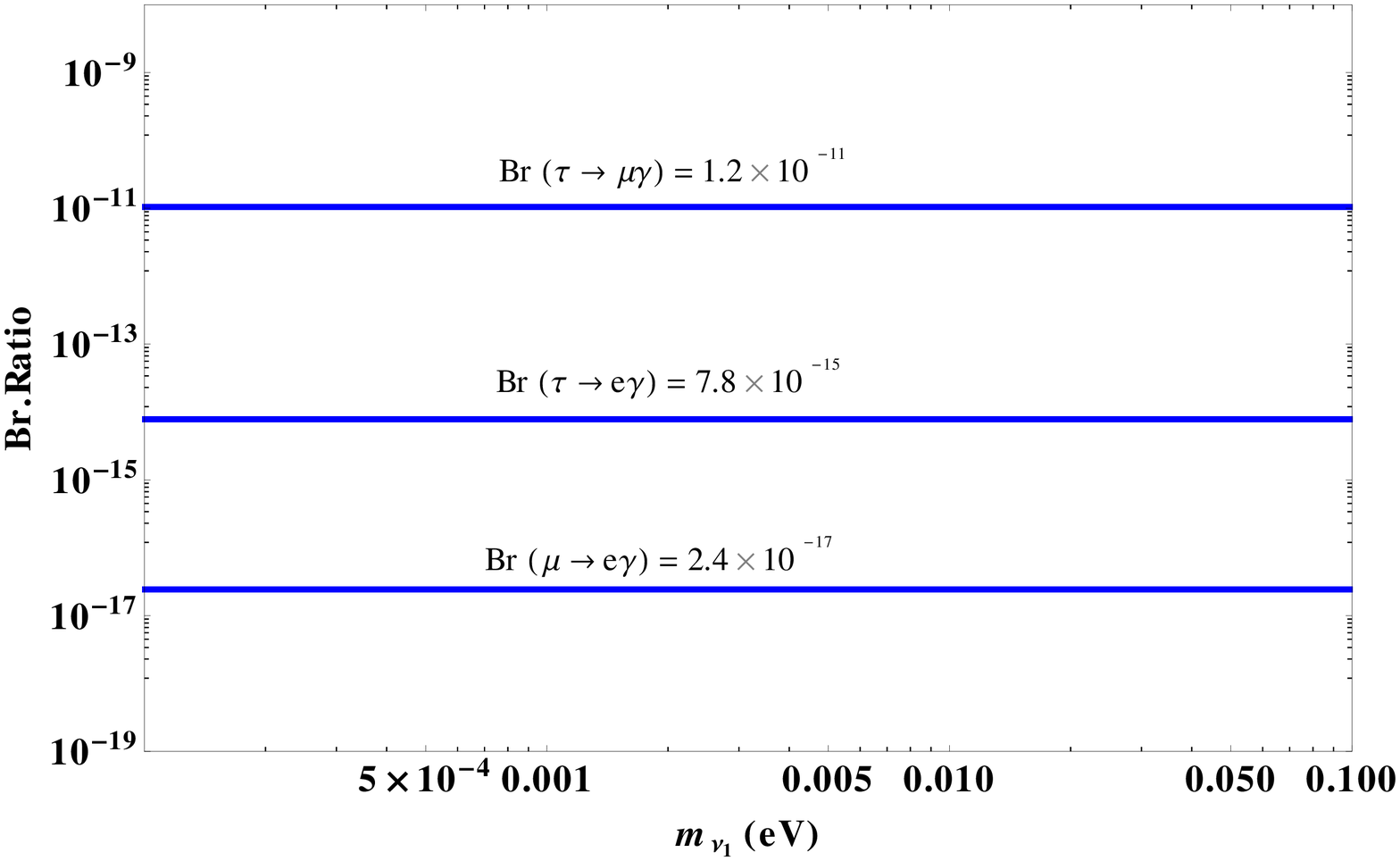}
 \caption{Variation of scalar triplet mass $M_{\Delta_L}$ (left-panel)
   and LFV branching ratio (right-panel) as a function of the lightest neutrino mass for different values of 
$\lambda$ in a $type-II$ seesaw dominant model where we have followed
   normal ordering. The three alomost
   horizontal lines represent the LFV branching ratios for $M=1.3$ TeV. }   
\label{Fig:bblfvt2}
\end{figure}

\begin{table}[htbp]
 \begin{tabular}{|c|c|c|c|c|}
  \hline
  $M_{1,2}$&$M_{3}$&$BR(\mu \to e\gamma)$&$BR(\tau
  \to e\gamma)$&$BR(\tau \to \mu\gamma)$\\
  (GeV)&(GeV)&&&\\ 
$1213.11$&$1213.11$&$6.43\times 10^{-17}$&$8.0\times10^{-16}$&$2.41\times 10^{-12}$ \\ \hline
$500$&$1280$&$1.2\times 10^{-16}$&$4.7\times10^{-15}$&$1.45\times 10^{-12}$\\ \hline
$100$&$1319.67$&$1.61\times 10^{-16}$&$6.04\times10^{-15}$&$1.80\times 10^{-12}$\\ \hline
$50$&$1545.31$&$9.0\times 10^{-15}$&$3.40\times10^{-14}$&$8.0\times 10^{-12}$ \\ \hline
\hline
 \end{tabular}
 \caption{
    Branching ratios for LFV decays $\mu \to e\gamma$, 
    $\tau \to e\gamma$, and $\tau \to \mu\gamma$ as function of
    $M_i(i=1,2,3)$}
 \label{tabcpv}
 \end{table}
 As a demonstration of solutions to the conceptual and practical issue
 of predicting experimentally accessible LFV decay branching ratios consistent with tiny
 neutrino masses in the dominant seesaw mechanisms ( in the event of type-I
 seesaw cancellation)  we present results
 in two specific  non-SUSY examples in SO(10):(i)type-II dominance \cite{bpn-mkp:2015},
 (ii) extended inverse seesaw dominance
 \cite{mkp-sahoo:NP:2015,
pp:2013,app:2013,pas:2014}. We have used the Dirac neutrino
 mass matrices and the fixed value of the matrix $\mu_S$ from the respective references.
Our predictions for branching ratios as a function of the lightest
neutrino mass are shown in in Fig.\ref{Fig:bblfvt2} for the type-II
dominance case. In this figure we have also shown 
variation of the LH triplet mass as expected from the type-II seesaw formula.
But inspite of the large value of the triplet mass that normally
predicts negligible LFV branching ratios, our model gives
experimentally accessible values.    

\begin{figure}[htb!]
\includegraphics[width=6cm,height=6cm]{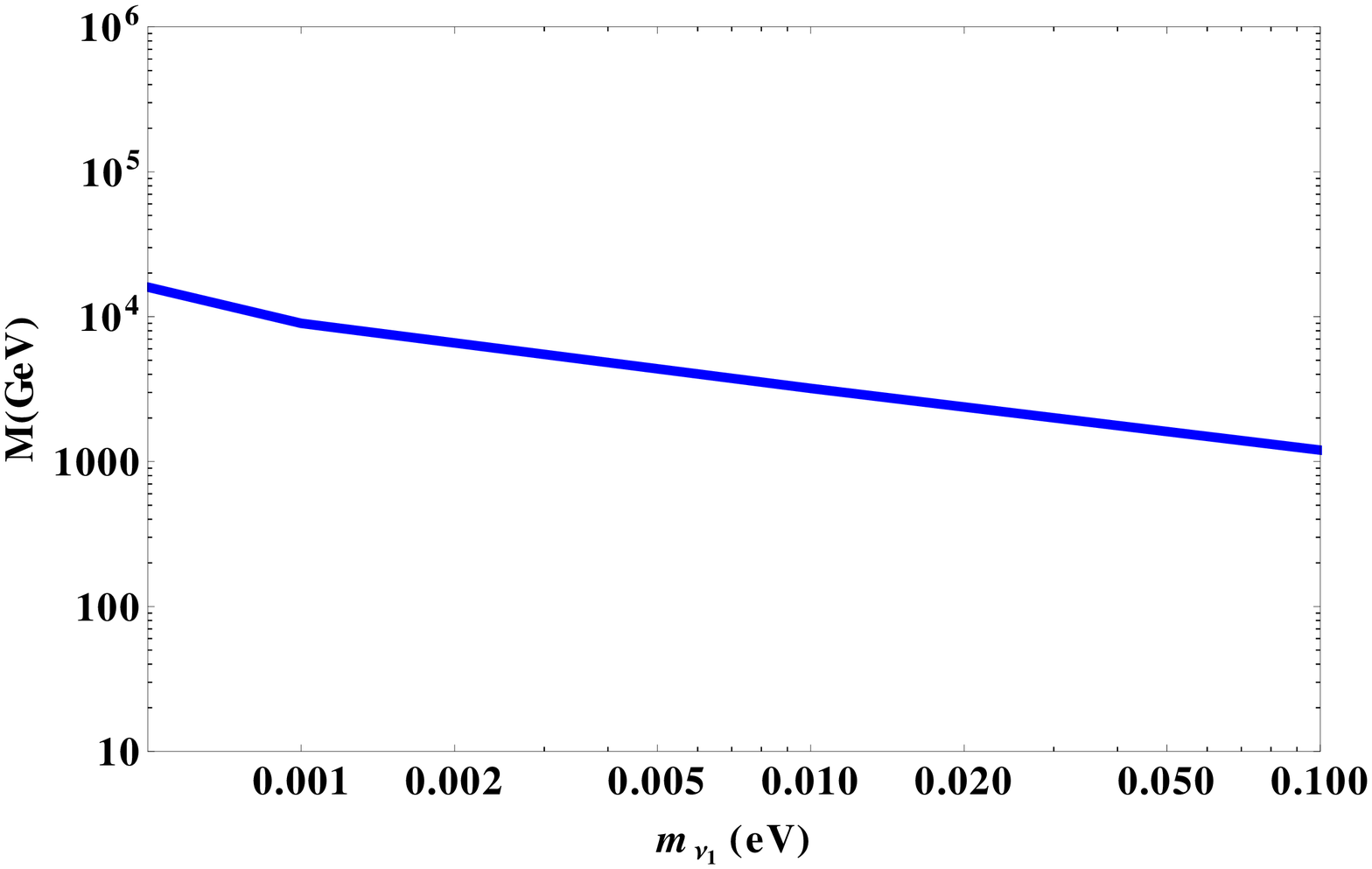}
\includegraphics[width=6cm,height=6cm]{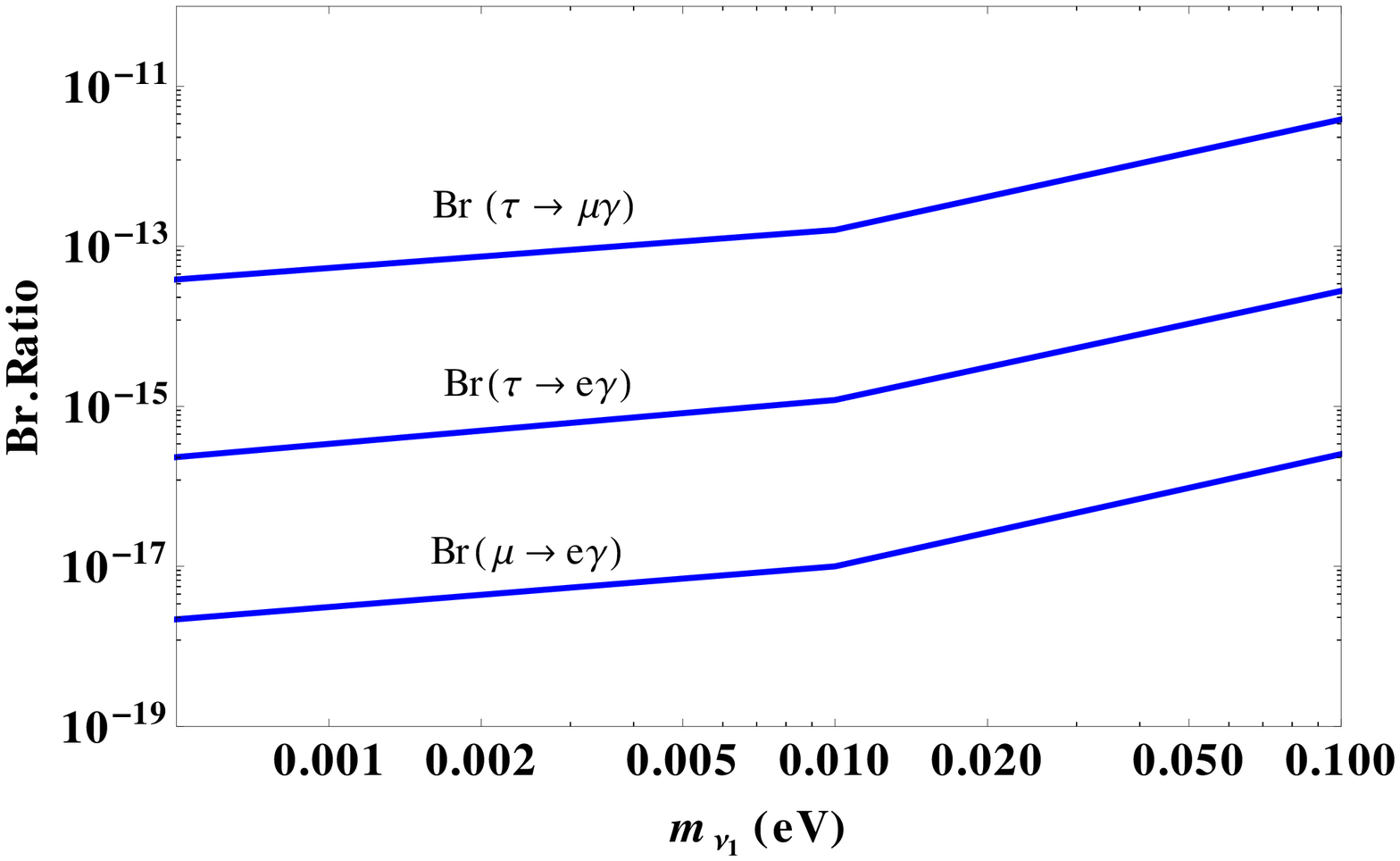}
 \caption{Same as Fig.\ref{Fig:bblfvt2} but for extended inverse
   seesaw dominance. The left panel represents variation of $N-S$
   mixing mass term $M$ as a function of lightest neutrino mass where
   we have followed normal ordering. The three lines in the right panel
    represent the LFV branching ratios.}  
\label{Fig:bblfvextnd}
\end{figure}

The corresponding results for the extended inverse seesaw case is
shown in Fig.\ref{Fig:bblfvextnd} for the degenerate values of the
$N-S$ mixing mass term $M=M_1=M_2=M_3=1.2$ TeV

We note that the predictions of LFV decay braching with classic
inverse seesaw \cite{Inverse} embedding in non-SUSY SO(10) as investigated in
\cite{ap:2012} with TeV scale Z' boson would be also similar to
Fig.\ref{Fig:bblfvextnd}. 

\subsection{CP-Violation due to Non-Unitarity}
The standard contribution to the CP violation is determined by the rephasing invariant 
$J_{\rm CP}$ associated with the Dirac phase $\delta_{CP}$ and matrix elements of the PMNS matrix 
$$J_{\rm CP}\equiv \text{Im}\left(U_{\alpha\, i}U_{\beta\, j} U^*_{\alpha\,j} U^*_{\beta\, j}\right) 
= \cos \theta_{12}\, \cos^2 \theta_{13}\, \cos \theta_{23}\, 
                      \sin \theta_{12}\, \sin \theta_{13}\, \sin \theta_{23}\, \sin \delta_{\rm CP}.$$ 
Because of the presence of non-unitarity effects 
, the leptonic CP-violation can be written as
\begin{eqnarray}
\mathcal{J}^{ij}_{\alpha \beta} =  \text{Im}\left(\mathcal{V}_{\alpha\, i} 
\mathcal{V}_{\beta\, j} \mathcal{V}^*_{\alpha\,j} \mathcal{V}^*_{\beta\, j}\right)
\simeq J_{\rm CP} + \Delta J^{ij}_{\alpha \beta}\,, 
\end{eqnarray}
where \cite{mkp-sahoo:NP:2015,bpn-mkp:2015,Antusch,psb-rnm:2010,ap:2012,app:2013,Ilakovac,Malinsky}
\begin{eqnarray}
\Delta J^{ij}_{\alpha \beta} = - \sum_{\rho=e, \mu, \tau} & &\text{Im} \bigg[  
                      \eta_{\alpha \rho}\, U_{\rho i}\, U_{\beta j}\, U^*_{\alpha j}\, U^*_{\beta i}  
                    + \eta_{\beta \rho}\, U_{\alpha i}\, U_{\rho j}\, U^*_{\alpha j}\, U^*_{\beta i} 
\nonumber \\  &&     + \eta^*_{\alpha \rho}\, U_{\alpha i}\, U_{\beta j}\, U^*_{\rho j}\, U^*_{\beta j} 
                    + \eta^*_{\beta \rho}\, U_{\alpha i}\, U_{\beta j}\, U^*_{\alpha j}\, U^*_{\rho j} \bigg]\,\label{deltaj} .
\end{eqnarray}
Model predictions for deviations from rephasing invariant matrix
defined in eq.(\ref{deltaj}) are
presented in Table \ref{tabcpv1} for different allowed values of
$M_i\equiv m_{R_i}$.
\begin{table}[htbp]
 \begin{tabular}{|c|c|c|c|c|c|c|}
  \hline
  $M_{1,2}$&$M_{3}$&$\Delta {\cj}^{12}_{e\mu}$&$\Delta
  {\cj}^{23}_{e\mu}$&$\Delta {\cj}^{23}_{\mu\tau}$&$\Delta
   {\cj}^{31}_{\mu\tau}$&$\Delta {\cj}^{12}_{\tau e}$\\
  (GeV)&(GeV)&&&&&\\ 
$1213.11$&$1213.11$&$-2.1\times10^{-6}$&$-2.4\times10^-{6}$&$1.4\times10^-{4}$&
$1.2\times10^{-4}$&$1.1\times 10^{-4}$ \\ \hline
$500$&$1280$&$-2.1\times10^{-6}$&$-2.4\times10^-{6}$&$1.4\times10^-{4}$&
$1.2\times10^{-4}$&$1.1\times 10^{-4}$ \\ \hline
$100$&$1319.67$&$-2.0\times10^{-6}$&$-3.0\times10^-{6}$&$1.3\times10^-{4}$&
$1.8\times10^{-4}$&$3.4\times 10^{-5}$ \\ \hline
$50$&$1545.31$&$-1.7\times10^{-6}$&$-1.98\times10^-{6}$&$1.1\times10^-{4}$&
$1.8\times10^{-4}$&$8.4\times 10^{-5}$ \\ \hline
\hline
 \end{tabular}
 \caption{Allowed deviations of the rephasing invariant matrix $J_{CP}$ as a result of
   weak leptonic CP-violation and non-unitarity effects defined in eq.(\ref{deltaj}) }
 \label{tabcpv1}
 \end{table}
 In order to visualise how far the predicted LFV decay branching ratios  are in
 agreement with neutrino mass values derived from the new seesaw
 mechanisms which survive after type-I cancellation, we have examined
 two cases as examples: (i)Type-II seesaw dominance
 \cite{bpn-mkp:2015} and (ii) extended seesaw dominance
 \cite{pp:2013,app:2013,pas:2014}. We have used the values of the
 Dirac neutrino mass matrix and the value of $\mu_S$ from these
 references.

\section{NEUTRINOLESS DOUBLE BETA DECAY}\label{sec.6}
\subsection{Double Beta Decay Predictions in the $W_R-W_R$ Channel}
The ongoing experiments on double beta decay without any conclusive results have led to a surge of investigations through different  theoretical predictions including LR gauge theories \cite{Rodejohann:2013,Tello,Deppisch:bb:rev12}. While the standard light neutrino exchange amplitude is $\propto G_F^2\sum_i U_{ei}^2m_i$ in the $W_L-W_L$
channel, the RH neutrino exchange amplitude in the $W_R-W_R$ channel is \cite{RNM:1986,Hirsch:1996,Tello-miha-nesti-gs-vissani:2011} $\propto G_F^2[\frac{M_W}{M_{W_R}}]^4\frac{1}{M_N}$. Thus the major  suppression factor $[{M_W}/{M_{W_R}}]^4 <10^{-4}$ apart from the  inverse proportionality factor  $M_N^{-1}$. The RH triplet exchange contribution  is also suppressed \cite{rnm-vergados:1981} and the  available experimental limit     
on the double beta decay rates has led to the lower bound on 
the doubly charged Higgs mass \cite{Tello}
\ba
M_{\Delta_R^{++}}&\ge&500\left(\frac{3.5 {\rm TeV}}{M_{W_R}}\right)^2 \nonumber\\
             &&\times{\left(\frac{M_N}{3 {\rm TeV}}\right)}^{1/2}.\label{rppm}  
\ea
Possible dominant contribution \cite{pp:2013} in the $W_L-W_R$ channel has been also suggested due to
what are known as $\eta-$ and $\lambda-$ diagrams \cite{Rodejohann:2013}.
It has been concluded \cite{RNM:rev16} that the lack of observation of double beta decay can not rule out TeV scale LR models.

\subsection{Double Beta Decay Predictions in the $W_L-W_L$ Channel}

\begin{figure}[htb!]
\includegraphics[width=8cm,height=8cm]{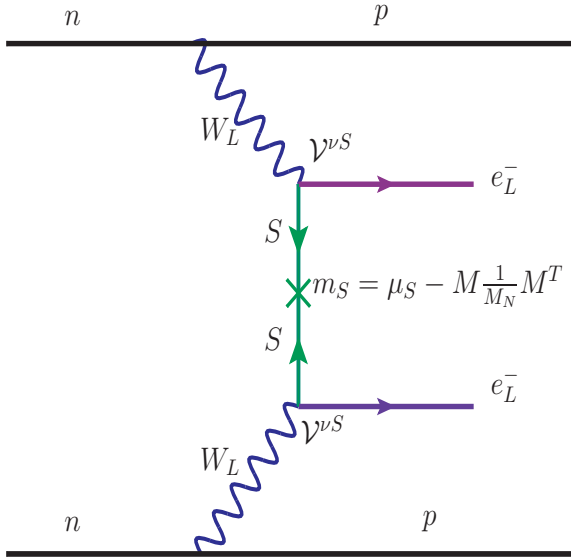}
 \caption{Feynman diagram for neutrinoless double beta decay  due to
   exchange of light sterile neutrino  $S$} .    
\label{Fig:bbflf2}
\end{figure}

In \cite{app:2013} it was  noted in the context of extended
gauged inverse seesaw mechanism in non-SUSY $SO(10)$ that a new and dominant
contribution to $0\nu\beta\beta$ decay exists through left-handed weak
charged currents in the $W_L-W_L$ channel via exchanges of fermion
singlets $S_i$. That the light sterile neutrinos may ha
ve a dominant
effect was also noted in the context of extensions of SM or LR models
 \cite{Mitra-gs-vissani:2012} where type-I contributions to neutrino
 mass was also included. As such this model \cite{Mitra-gs-vissani:2012}    did not use cancellation mechanism for the type-I seesaw contributions.  The sterile fermions mix
quite prominently through Dirac neutrino mass matrix which is well
known to be of order of up quark mass matrix. As already noted the sterile neutrinos  acquire
  Majorana masses $m_S\simeq \mu_S
-M\frac{1}{M_N}M^T\simeq -M\frac{1}{M_N}M^T$.  The Feynman diagram for
 double beta decay due to light sterile neutrino exchange is shown in Fig.\ref{Fig:bbflf2}. 

  The 
$(ei)$ element of the $\nu-S$ mixing matrix is
\be
 {\mathcal{V}}^{\nu S}_{ei} = (\frac{M_D}{M})_{ei},\label{vnus}   
\ee
where the Dirac neutrino mass matrix $M_D$ has been given in different
$SO(10)$ models from fits to the charged fermion masses at the GUT
scale and then running it down to the TeV scale in non-SUSY SO(10)
models \cite{mkp-sahoo:NP:2015,bpn-mkp:2015,
ap:2012,app:2013,pas:2014}.
For example the Dirac neutrino mass matrix $M_D$ in the model\cite{bpn-mkp:2015} is
\bea
M_D({\rm GeV})= 
\footnotesize\begin{pmatrix}
0.014&0.04-0.01i&0.109-0.3i\\
0.04+0.01i&0.35&2.6+0.0007i\\
0.1+0.3i&2.6-0.0007i&79.20
\end{pmatrix}.\nonumber\\
\label{MDatMR0} 
\eea 
 The elements of the matrix $M$ treated as diagonal is determined from
constraints on LFV decays and
the diagonal elements are estimated using the non-unitarity equation
as discussed in the previous section.
We derive the relevant elements of the mixing matrix
$\mathcal{V}^{\nu S}$ using the structures of the
the Dirac neutrino mass matrix $M_D$ given in eq.(\ref{MDatMR0}) and
values of the diagonal elements of $M=(M_1, M_2, M_3)$ satisfying the
non-unitarity constraint in eq.(\ref{costr}). The eigen values of the
$S-$ fermion mass matrix are estimated for different cases using the
structures of the  
 RH Majorana neutrino
mass matrices \cite{bpn-mkp:2015,
pp:2013,app:2013,pas:2014} and allowed values
of $M_i$ satisfying the non-unitarity constraints through the formula
$M_S=-M\frac{1}{M_N}M^T$.

Different particle exchange contributions for $0\nu 2\beta$ decay are 
changed by the chirality of the hadronic currents involved in the
nuclear mass matrix element. In this model only the LH currents are
significant or dominant while the RH current effects are
negligible. The inverse half-life has been estimated  with proper
normalization factors by taking into account the effects of nuclear
matrix elements \cite{Rodejohann:2013,
RNM:1986,Doi:1993,Vergados:1996} 
\begin{eqnarray}
  \left[T_{1/2}^{0\nu}\right]^{-1} &\simeq &
  G^{0\nu}_{01}|\frac{{\cal M}^{0\nu}_\nu}{m_e}|^2|({\large\bf
    m}^{ee}_{\nu} +{\large\bf m}^{ee}_{S}+{\large\bf m}^{ee}_{N})|^2 \nonumber\\
&\simeq& K_{0\nu}|({\large\bf
    m}^{ee}_{\nu} +{\large\bf m}^{ee}_{S})|^2.
  \label{eq:invhalflife_simp}
\end{eqnarray}
The larger value of RH neutrino masses make negligible contribution to the $0\nu\beta\beta$ decay amplitude.
In the above equation  $G^{0\nu}_{01}= 0.686\times10^{-14}$ yrs$^{-1}$, ${\cal
  M}^{0\nu}_{\nu} = 2.58-6.64$, $K_{0\nu}= 1.57\times 10^{-25}$ {yrs}$^{-1}
{\rm eV}^{-2}$ and the two effective mass parameters are
\begin{eqnarray}
&& 
{\large \bf  m}^{\rm ee}_{\nu} =\sum^{}_{i} \left(\mathcal{V}^{\nu \nu}_{e\,i}\right)^2\, {m_{\nu}}, \\ 
&&
{\large \bf  m}^{\rm ee}_{S} = \sum^{}_{i} \left(\mathcal{V}^{\nu S}_{e\,i}\right)^2\, \frac{|p|^2}{m_{S_i}},\\
&&
{\large \bf  m}^{\rm ee}_{N} = \sum^{}_{i} \left(\mathcal{V}^{\nu N}_{e\,i}\right)^2\, \frac{|p|^2}{m_{N_i}}.
\label{effmassparanus} 
\end{eqnarray}
Here $m_{S_i}$ is the eigen value of the $S-$ fermion mass 
\be
 M_S=-M\frac{1}{M_N}M^T.\label{smatrix}
\ee

\subsection{Singlet Fermion  Exchange Dominated Half-Life} 
\begin{figure}[htbp]
\begin{center}
\includegraphics [width=6cm,height=6cm]{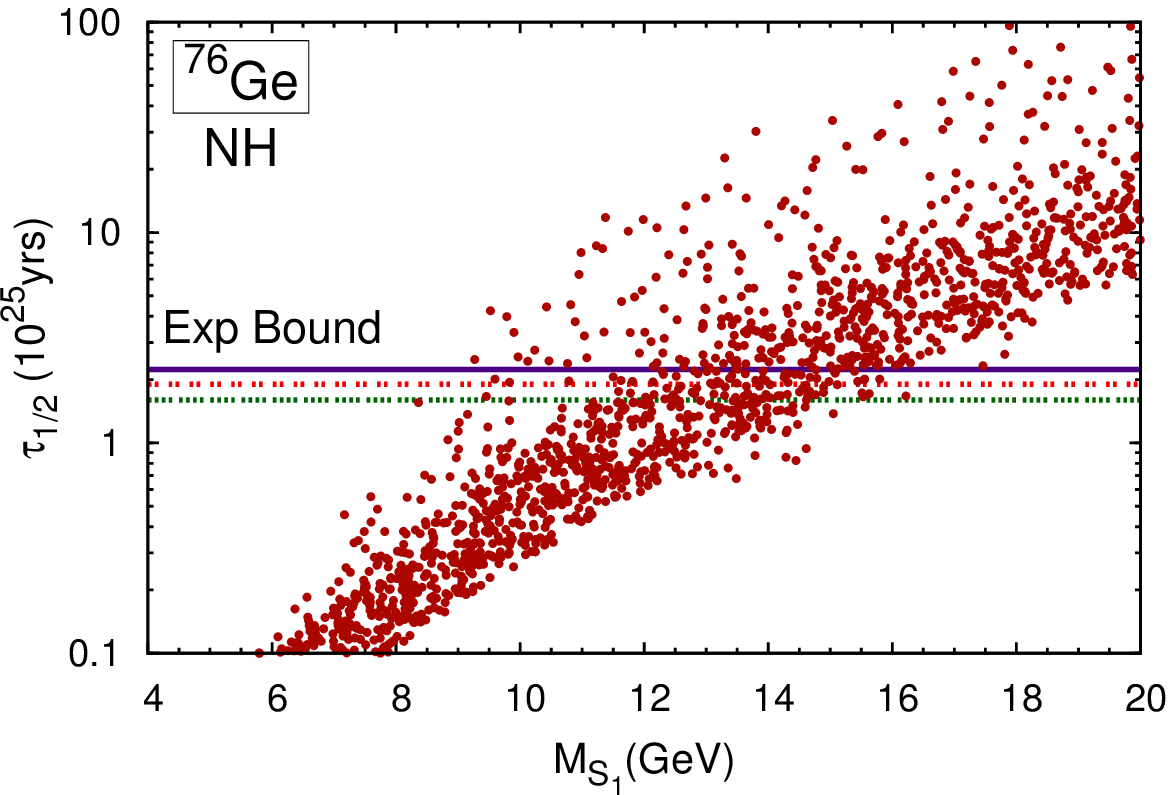}
\includegraphics [width=6cm,height=6cm]{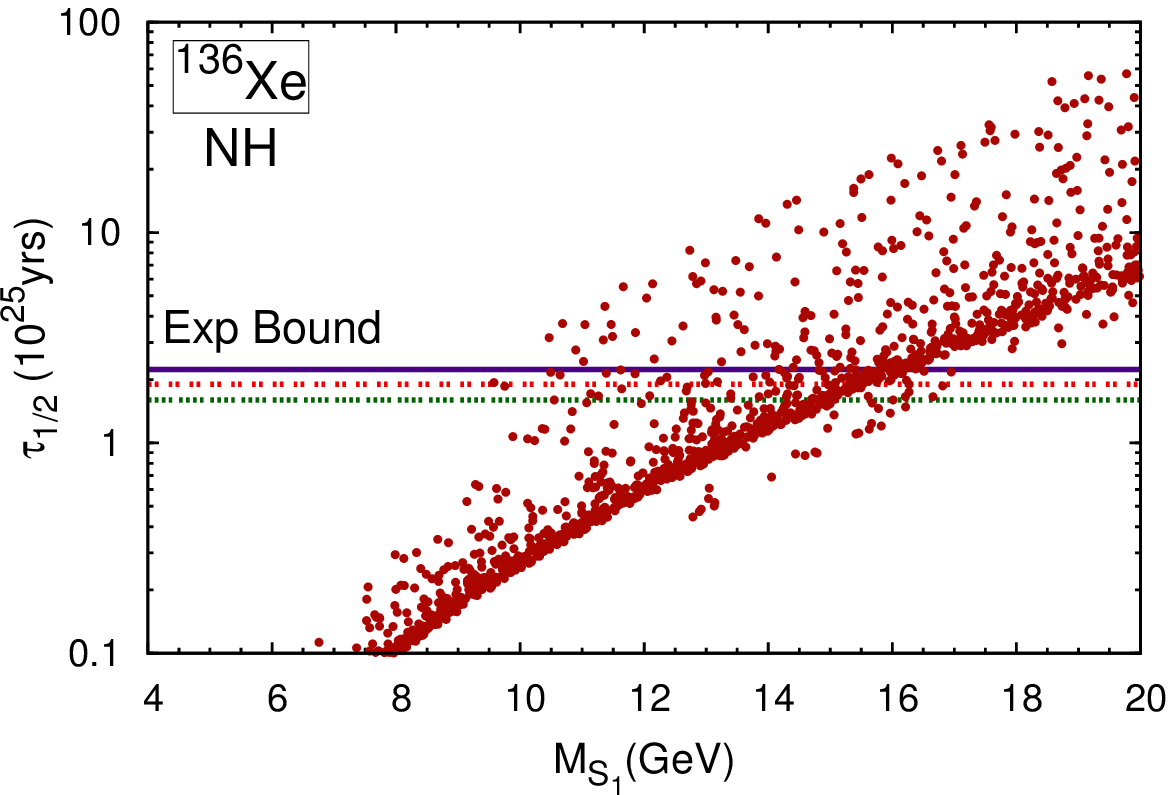}
\end{center} 
\caption{Scattered plots for half-life prediction of double beta decay
  as a function of neutral singlet fermion mass in the
  case of normal hierarchy (NH) of light neutrino masses
  from ${}^{76}Ge$ nucleus (left-panel) and from ${}^{136}Xe$ nucleus
  {right panel}.}
 \label{Fig:bb1}
 \end{figure}

\begin{figure}[htbp]
\begin{center}
\includegraphics [width=6cm,height=6cm]{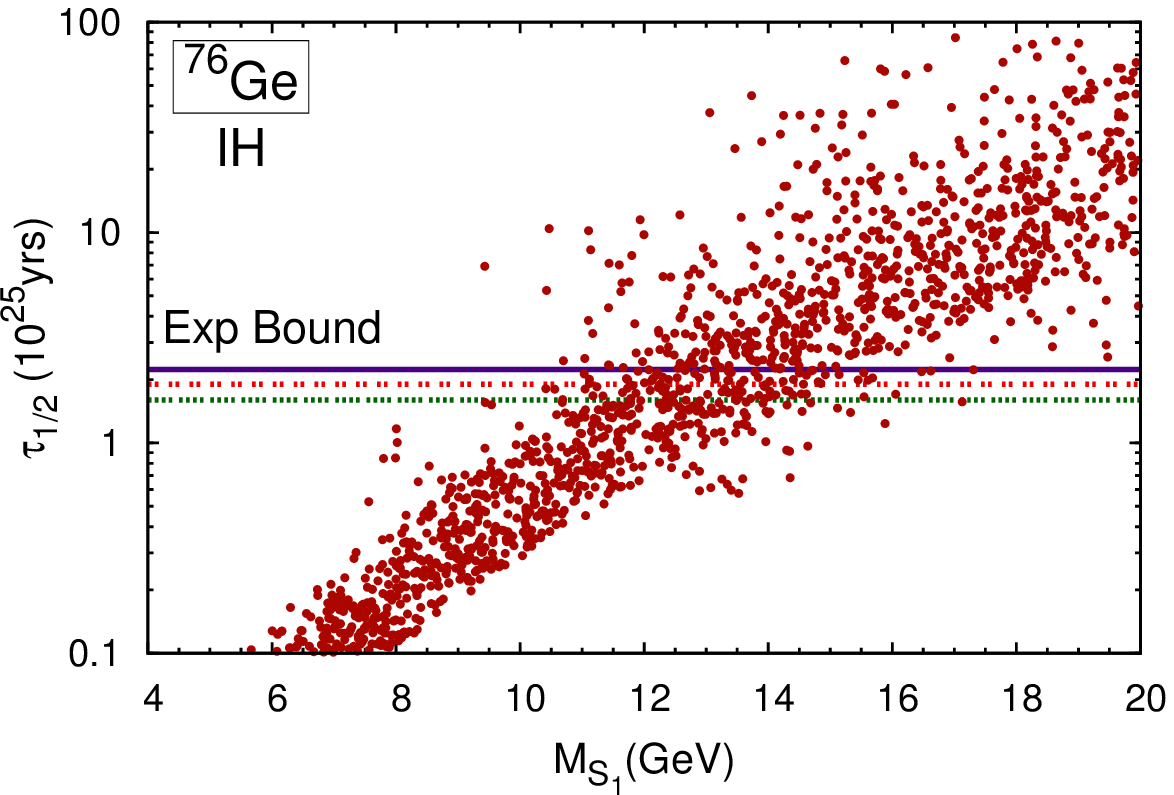}
\includegraphics [width=6cm,height=6cm]{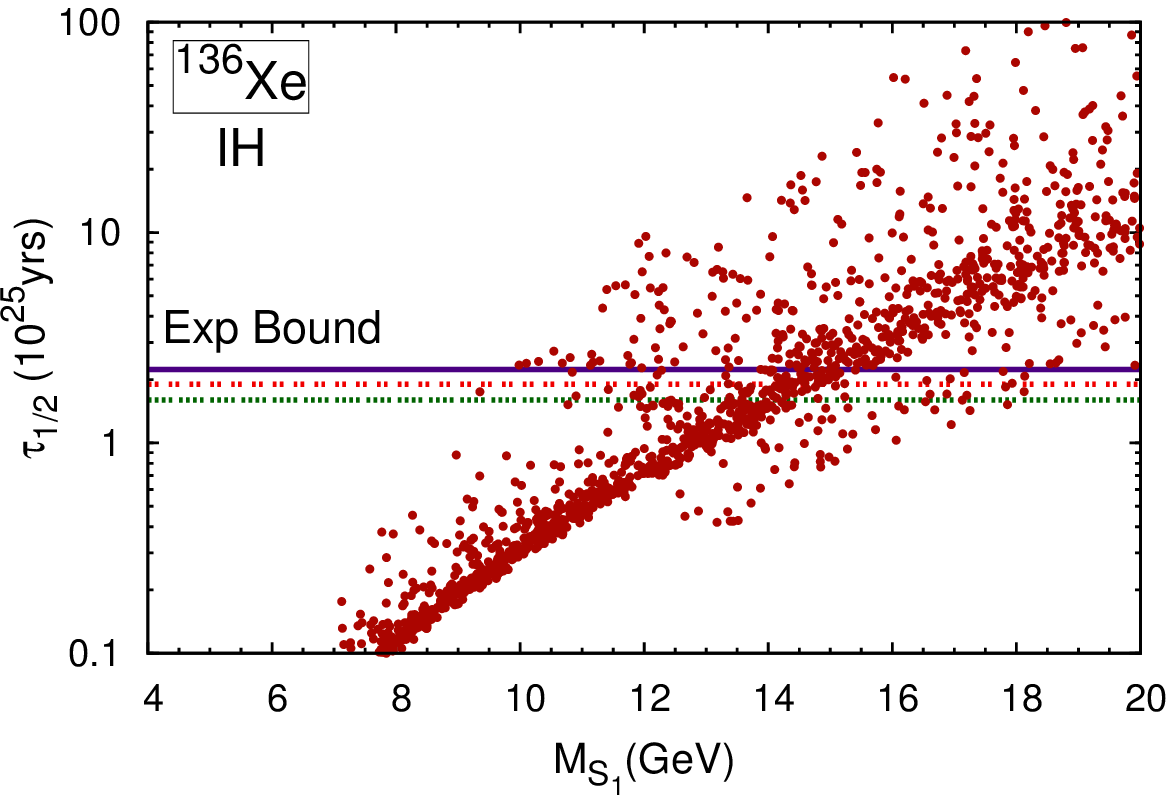}
\end{center} 
\caption{Same as Fig.\ref{Fig:bb1} but for inverted hierarchy (IH) of
  light neutrino masses. }
 \label{Fig:bb2}
 \end{figure}

\begin{figure}[htbp]
\begin{center}
\includegraphics[width=6cm,height=6cm]{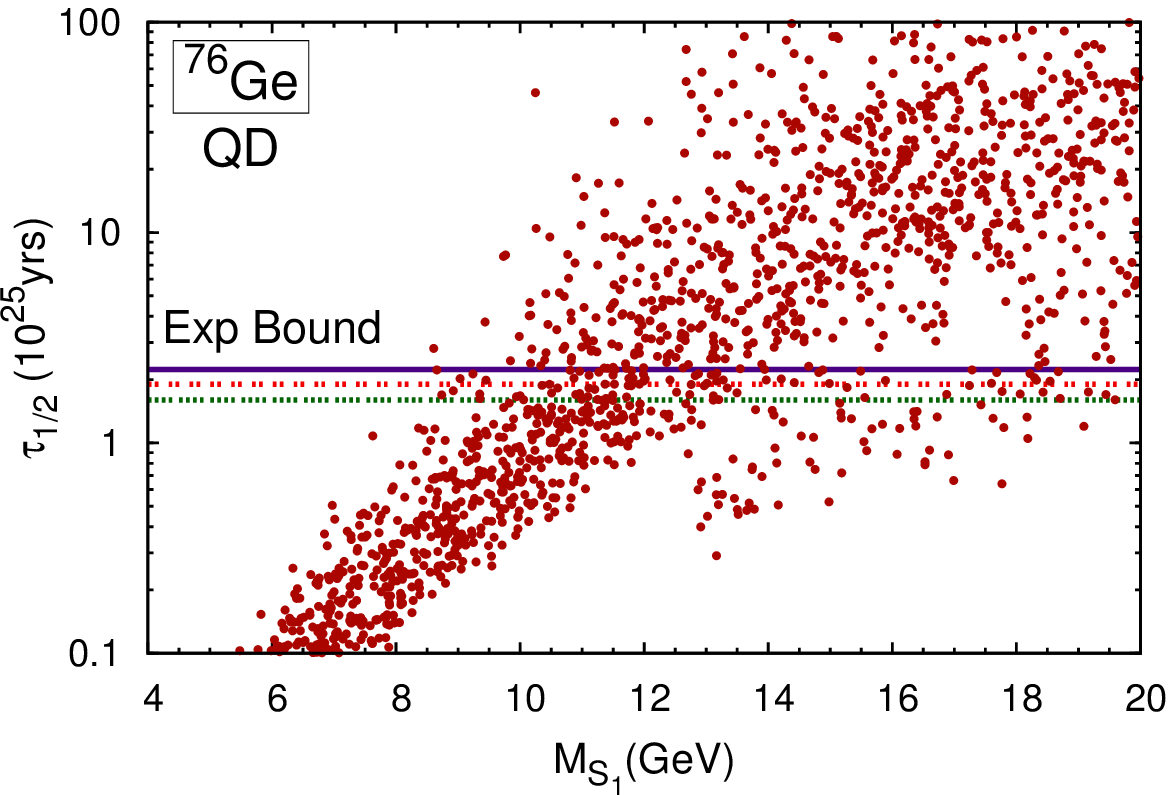}
\includegraphics [width=6cm,height=6cm]{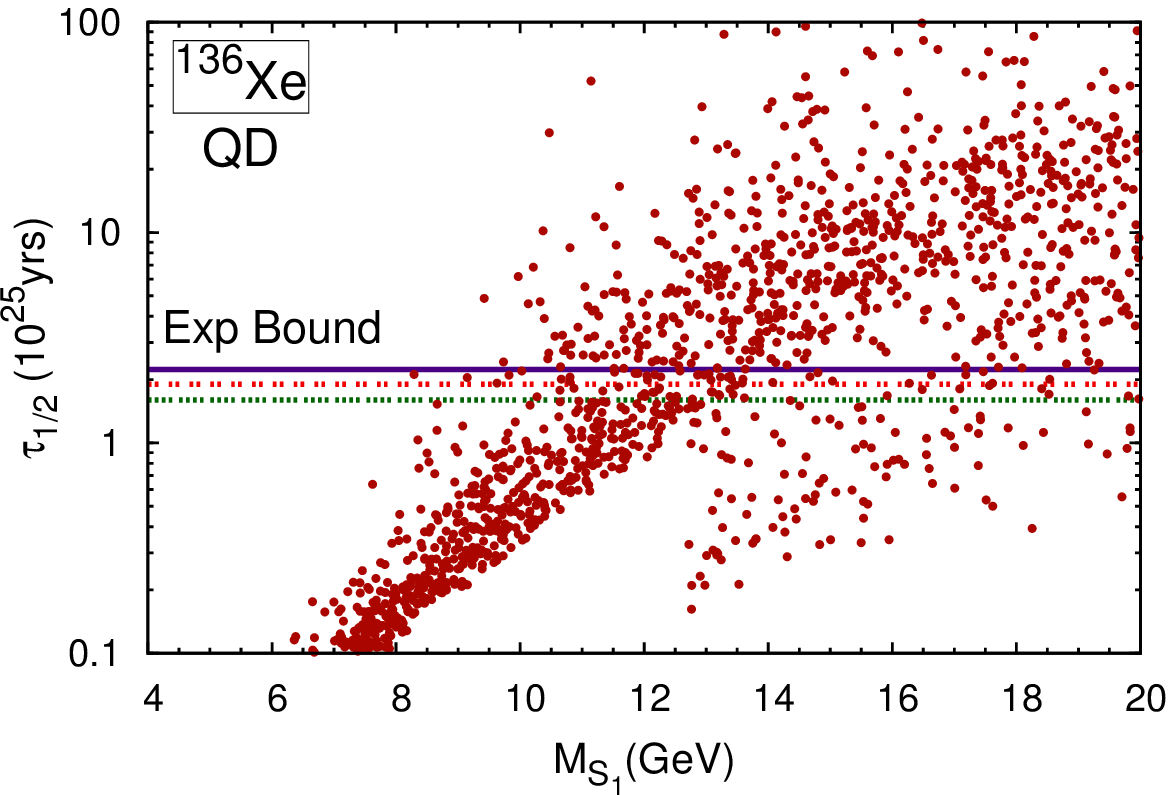}
\end{center} 
\caption{Same as Fig.\ref{Fig:bb1} but for quasi degenerate (QD) type
  light neutrino masses. }
 \label{Fig:bb3}
 \end{figure}

The nonstandard contributions to half life of double beta decay as a function
of sterile neutrino mass has been discussed  in
\cite{mkp-sahoo:NP:2015,bpn-mkp:2015,
ap:2012,app:2013,pas:2014}. The models predict $\nu-S$ mixing and
sterile neutrino masses which have been used to predict the half-life
in the case of different hierarchies of light neutrino masses:NH, IH,
and QD. Using the estimations of ref.\cite{pas:2014} which are also
applicable to models of
refs.\cite{mkp-sahoo:NP:2015,bpn-mkp:2015,pp:2013,app:2013}, scattered plots
for the predicted half-life are
shown in Fig.\ref{Fig:bb1},  Fig.\ref{Fig:bb2},  and  
Fig.\ref{Fig:bb3} for the three types of neutrino mass hierarchies.

\begin{figure}[htbp]
\begin{center}
\includegraphics[scale=0.8]{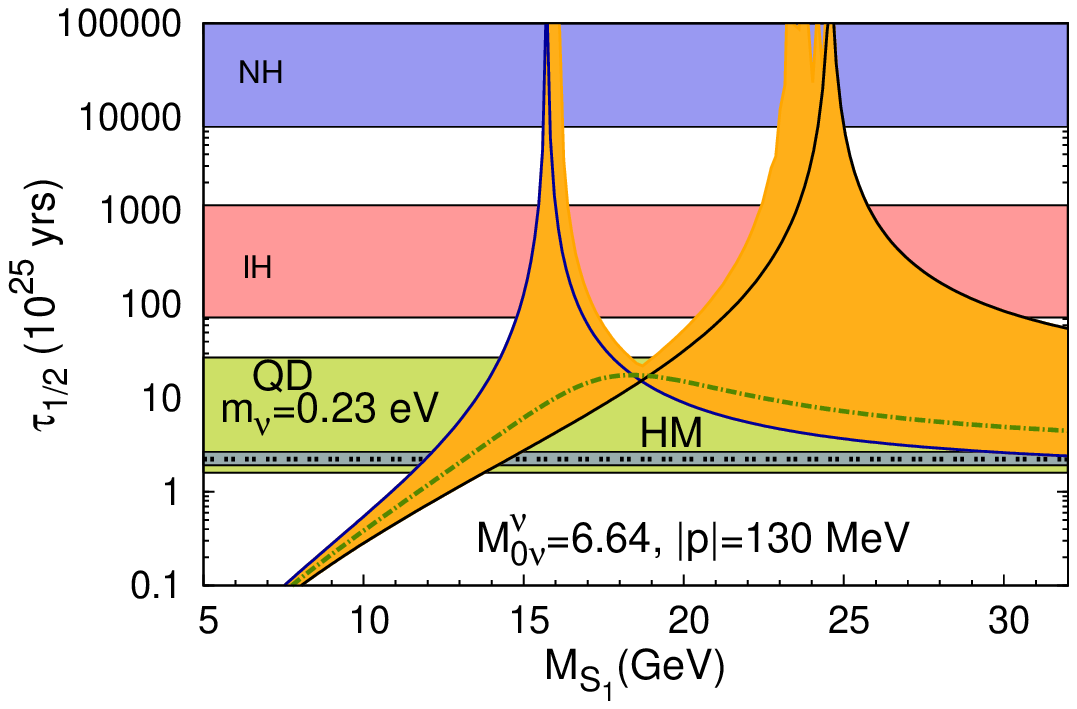}
\end{center} 
\caption{Prediction of half-life for double beta decay as a function
  of sterile neutrino mass in the case of QD type mass hierarchy with
  the common mass parameter $m_0=0.23$ eV. The peaks correspond to
  cancellation between light neutrino and sterile neutrino exchange
  amplitudes when Majorana CP phases of sterile neutrino is
  ignored. The dotted line shows the absence of peaks when CP phases
  are included \cite{pas:2014} }
 \label{Fig:bb4}
 \end{figure}

  In Fig.\ref{Fig:bb4} we show predictions in the QD case excluding
  and including the CP phases of Majorana type sterile neutrinos. The
  cancellations between light neutrino exchange amplitude and the
  sterile neutrino exchange amplitude is shown by the two peaks. When
  CP phases associated with the Majorana type sterile neutrino mass eigen value(s) are included the peaks are smoothened as shown by dotted lines\cite{pas:2014}. 

\begin{figure}[htbp]
\begin{center}
\includegraphics[scale=0.8]{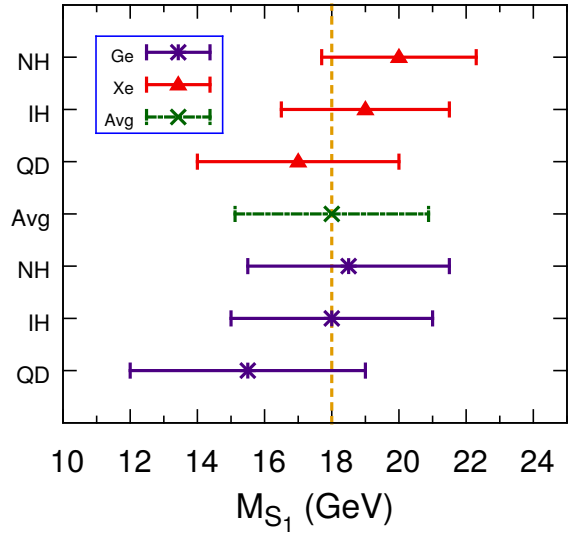}
\end{center} 
\caption{Prediction of light sterile neutrino mass from the saturation
of experimental decay rates of ongoing searches for different active
neutrino mass hierarchies. The horizontal green line indicates the
average value of all results. }
 \label{Fig:bb5}
 \end{figure}

In Fig.\ref{Fig:bb5} estimations on the lightest sterile neutrino
masses are predicted which saturate the current experimental limit on
the observed double beta decay half life using Ge$-76$ and Xe$-136$
nuclei for three different light neutrino mass hierarchies in each
case. The uncertainties in the predicted masses correspond to the
existing uncertainty in the neutrino virtuality momentum $|p|=120 -
200$ MeV. The green horizontal line represents the average value 
\begin{equation}
{\hat M}_{S_1}= 18 \pm 4~~~~{\rm GeV}, \label{bbsmass} 
\end{equation}
of the lightest sterile neutrino mass determined from double beta decay 
experimental bound \cite{pas:2014}. Lower values of this mass has been
obtained using light neutrino assisted type-II seesaw dominance \cite{bpn-mkp:2015}.

These predictions suggest that sterile neutrino exchange contribution
dominates the double beta decay rate even when the light neutrino
masses have NH or IH type of mass hierarchies. To predict double beta
decay saturating the current experimental bounds, it is not necessary
that light neutrinos should be quasi-degenerate in mass.
On the other hand if double-beta decay is not found with half-life
close to the current limits, then the solutions with light sterile
neutrino masses in the range $\sim (2-15)$ GeV are ruled out, but the
model with larger mass eigen values easily survives. 

\section{LEPTOGENESIS IN EXTENDED MSSM AND  SUSY SO(10)}\label{sec.7}

In the conventional type-I seesaw based leptogenesis models where heavy RH
neutrino decays give rise to the desired lepton asymmetry \cite{Fuku-Yana:1986}, the
Davidson-Ibarra bound \cite{DI:2002,Ham-gs:2004} imposes a lower limit on the scale of
leptogenesis, $M_{N_1} > 4.5\times 10^9$ GeV \cite{DI:2002,Ham-gs:2004}. This also suggests the
lower bound for the reheating temperature after inflation, $T_{RH}\ge
10^9$ GeV, that would  lead to
overproduction of gravitinos severely affecting the relic abundance of
light nuclei since the acceptable limit has been set as  $T_{RH}\le 10^7$ GeV 
 \cite{Khlopov:1984}. Several attempts have been made to
evade the gravitino constraint on leptogenesis where sterile neutrino
assisted results are interesting. Obviously gravitino constraint is
satisfied in models with TeV scale resonant leptogenesis
\cite{SUSY-Res}. Also in the singlet fermion extended SUSY SO(10)
where RH neutrinos are heavy pseudo Dirac neutrinos and neutrino mass
formula is through inverse seesaw \cite{Inverse} , there is no
problem due to gravitino constraint
\cite{psb-rnm:2010,Blanchet:2010}. We discuss the cases where all the
three types of neutrinos are Majorana fermions.
\subsection{Leptogenesis with Extended Seesaw Dominance}
With extended seesaw realisation of leptogenesis in two types of SUSY models have
been investigated under gravitino constraint:(i)MSSM
extension with fermion singlets
\cite{Kang-Kim:2006,Cheon:2007,Cheon:2009,Kang-Patra:2014}, and (ii)
Singlet extension of  SO(10) with
intermediate scale $G_{2213}$ gauge symmetry. 
\cite{Majee:PLB:2007,mkp-arc:2010}. We discuss their salient features.
\subsubsection{MSSM Extension with Fermion Singlets}
The Dirac neutrino mass matrix is identified with the charged lepton
mass matrix in this model where MSSM is extended with the addition of heavy RH neutrinos $N_i$
as well as additional singlets $S_i$ \cite{Kang-Kim:2006}, one for
each generation. As already explained in the limit $M_N > M >>M_D, \mu_S$
   extended seesaw
formula, which is the same as the inverse seesaw formula for active
neutrino mass. In this case resonant leptogenesis is 
implemented via 
quasi-degenerate  RH neutrino decays at the TeV scale. It is well known that
such resonant leptogenesis scenario with $M_{N_1}\sim M_{N_2} \sim 1$
TeV implemented through canonical seesaw, needs a very small mass
splitting between the RH neutrinos
$\frac{(M_{N_2}-M_{N_1})}{(M_{N_2}+M_{N_1})}\sim 10^{-6}$. With the
 the tension arising out of fitting the neutrino oscillation data being
transferred from type-I seesaw to extended seesaw in the presence of 
additional sterile fermions, it is not unimaginable that this
fine tuning associated with very tiny RH neutrino mass splitting could be
adequately alleviated.
 The fermion singlets $S_i$
give rise to a new self-energy contribution and using this successful resonant
leptogenesis has been found to be possible  for a much larger  mass ratio
$\frac{M_{N_2}}{M_{N_1}}\sim 10$. Possibilities of $\sim 100$ MeV to
  $\sim 10$ GeV 
  mass range for  light sterile neutrinos have been pointed out.  
In a separate analysis the possibility of singlet Majorana fermion
 or singlet scalar as candidates of dark matter has been pointed out
 \cite{Cheon:2007}. Realisation of doubly coexisting dark matter
 candidates in the context of extended seesaw framework has been
 pointed out \cite{Cheon:2009}.   
\subsubsection{Leptogenesis in SUSY SO(10)}
In non-SUSY minimal LR models where $M_D$ is similar to charged lepton
mass matrix successful leptogenesis emerges with intermediate scale
hierarchical RH neutrino masses \cite{Babu-Bachri-Aissoui:2006}.
In SUSY SO(10) the underlying quark-lepton unification forces the
Dirac neutrino mass to be similar to the up-quark mass matrix. This
pushes the type-I seesaw scale closer to the GUT scale, $M_R\ge
10^{14}$ GeV and rules out the possibility of low scale $W_R$ bosons
accessible to accelerator searches in foreseeable future unless the
canonical seesaw ansatz is given up, for example, in favour of inverse
seesaw with TeV scale pseudo Dirac neutrinos and $W_R$ bosons
\cite{psb-rnm:2010,Blanchet:2010}. With heavy right-handed Majorana
neutrinos and GUT-scale LR breaking scale, successful leptogenesis has
been implemented in realistic SUSY
SO(10) \cite{Ji:2007}. With the help of an effective ${\rm dim}.5$ 
operator ansatz  which originates from renormalisable interactions at
GUT-PlancK scale  in SUSY SO(10) ( without using
${126}_H$) both  thermal and non-thermal leptogenesis
\cite{Kumekawa:1994,Jannernot:2000} have been discussed with heavy
hierarchical RH neutrino of masses
\cite{Pati-lepto:2003}. Possible solutions to the allowed
parameterspace to evade gravitino constraint have been also discussed
in this work. 

Apart from the models with resonant
leptogenesis,  possibility of
leptogenesis under gravitino constraint in SUSY SO(10) has been
realised with hierarchical RH neutrinos assisted by sterile
neutrinos. As already noted above, in these cases the
extended seesaw formula controls the neutrino mass as a result of
cancellation of type-I seesaw contribution.  Gauge
coupling unification in these SO(10) models requires the $G_{2213}$ symmetry to
occur at the intermediate scale in the renormalizable model
\cite{mkp-arc:2010}. A common feature of both these models \cite{Majee:PLB:2007,mkp-arc:2010} is the generation of
lepton asymmetry through the decay of hierarchical sterile neutrinos
through their respective mixings with heavier RH neutrinos which are
also hierarchical. 
 
An important and specific advantage of heavy gauge-singlet neutrino decay to
achieve leptonic CP asymmetry is the following:
The singlet neutrino of mass $\sim 10^5$ GeV which
decays to $l\phi$ though its mixing with RH neutrino of mass $\sim
10^{10}$ GeV has a small mixing angle $\sim 10^{-5}$. This small mixing
ensures out-of-equilibrium condition by making the decay rate smaller
than the Hubble expansion rate in arriving at CP asymmetry at
lower temperatures $\sim 300$ GeV.

\section{SINGLET FERMION ASSISTED LEPTOGENESIS IN NON-SUSY SO(10)}\label{sec.8}

\begin{figure}[htbp]
\begin{center}
\includegraphics[width=4cm,height=4cm]{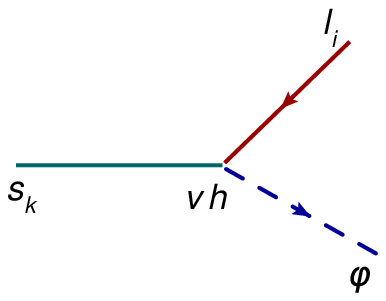}
\includegraphics[width=4cm,height=4cm]{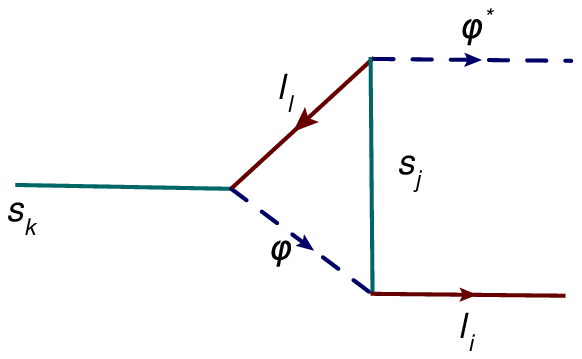}
\includegraphics[width=4cm,height=4cm]{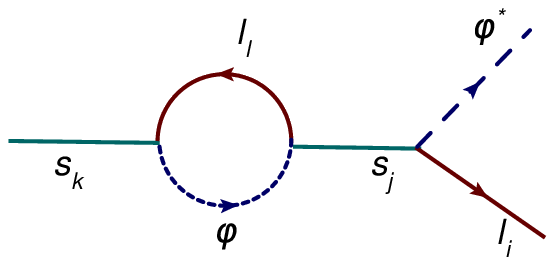}
\caption{Tree and one-loop diagrams for the $S_{k}$ decay contributing to the CP-asymmetry. All fermion-Higgs couplings in the diagrams are of the
form $Vh$ where $h= N-l-\Phi$ Yukawa coupling and $V\simeq M/M_N$.}
\label{fig:Fvertex}
\end{center}
 \end{figure}

An extensive review of thermal leptogenesis with reference to LFV is available in
\cite{Molinaro:2010}. With the neutrino mass following a modified
 type-I seesaw at a scale $\ge 10^8$ GeV, thermal leptogenesis has
 been investigated in ref.  
It is well known that TeV scale RH neutrinos can participate in
resonant leptogenesis contributing to enhanced generation of leptonic
CP-asymmetry that is central to generation of baryon asymmetry of the
universe via sphaleron interactions. Here we briefly discuss a recent
work where quasi-degenerate sterile neutrinos at the TeV scale in
non-SUSY SO(10) have been shown to achieve resonant leptogenesis
through their decays. The Feynman diagrams at the tree level and with
vertex and self energy corrections are shown in Fig.\ref{fig:Fvertex}.

The fermion-Higgs coupling in all the diagrams is
 $Vh$ instead of the standard Higgs-Yukawa coupling 
$h=M_D/ V_{\rm  wk}$ where  ${\mathcal V}\simeq{M/M_N}$,
$M_D$ is given in eq.(\ref{MDatMR0}), and $V_{\rm  wk}\simeq 174$ GeV.  
Denoting the mass eigen value of a sterile neutrino by ${\hat m}_{S_k}
(k=1, 2, 3)$,
for computation of  baryon asymmetry $Y_B$ of the Universe with a  washout
factor $K_k$, we utilise the ansatz
\cite{Pilaftsis-Underwood:2004} 
\bea
Y_B &\simeq& \frac{\varepsilon_{S_k}}{200 K_k},\nonumber\\ 
K_k &=&\frac{\Gamma_{S_k}}{H({\hat m}_{S_k})}, \label{bau}  
\eea
$H({\hat m}_{S_k})$ being the Hubble parameter at temperature ${\hat m}_{S_k}$.
Defining 
\be
\delta_i=\frac{|{\hat m}_{S_i}-{\hat m}_{S_j}|}{\Gamma_{S_i}}(i\neq
j), 
\ee
the depleted washout factor is \cite{Hambye-wash:2010}
\be
K_i^{\rm eff}\simeq \delta_i^2 K_i.\label{keff}
\ee 
Here we discuss two cases for the sterile neutrino contribution towards leptogenesis and baryon asymmetry:
(a)  ${\hat m}_{s_1}$ is light, ${\hat m}_{s_2}$ and
${\hat m}_{s_3}$ are quasi-degenerate;
(b) ${\hat m}_{s_2}$ is light, ${\hat m}_{s_1}$ and  ${\hat m}_{s_3}$ are quasi-degenerate.\\
\noindent{\bf{Case (a). ${\hat m}_{s_1}$  light, ${\hat m}_{s_2}$ and  ${\hat m}_{s_3}$ heavy and quasi-degenerate.}}\\
Using an allowed interesting region of the parameter
space $M \simeq {\rm diag.} (146,3500,3500)$ GeV,
  $V_R=10^4$ GeV, and $M_N=fV_R$  we get
\bea 
{\hat m}_{S_i} &=& {\rm diag.} (1.0, 595.864.., 595.864..){\rm
  GeV}.\label{msi1}
\eea 
leading to
 $K_2= 2.7\times 10^7$.  Using $({\hat m}_{S_2}-{\hat m}_{S_3})\simeq
2\times 10^{-7}$ GeV, we obtain
\bea
\varepsilon_{S_2}&=& 0.824,\nonumber\\
Y_B&=& 1.5\times 10^{-10}.\label{bau2f1}
\eea

\par\noindent{\bf{Case (b) ${\hat m}_{s_2}$ light,  ${\hat m}_{s_1}$ and  ${\hat m}_{s_3}$ heavy
    and quasi-degenerate.}}\\
 
Choosing another allowed region of the parameter space $M \simeq {\rm
  diag.} (3200,146,3200)$ GeV, similarly we get
 \bea 
{\hat m}_{S_i} &=& {\rm diag.} (500.567.. ,1.0, 500.567..){\rm
  GeV},\label{msi2}
\eea
leading to
 $K_1= 4\times 10^6$.  using $({\hat m}_{S_1}-{\hat m}_{S_3})\simeq
7\times 10^{-5}$ GeV, we obtain
\bea
\varepsilon_{S_1}&=& 0.682,\nonumber\\
Y_B&=& 4\times 10^{-10}.\label{bau2f2}
\eea

In Case (a) with  ${\hat m}_{S_1}\sim {\mathcal O}(1)$ GeV , the lightest sterile
 neutrino acts as the most
 dominant source of  $0\nu\beta\beta$ decay whereas the heavy quasi-degenerate
pair of sterile neutrinos $S_2$ and $S_3$ mediate resonant
leptogenesis. Similarly in the alternative scenario  of Case (b) with ${\hat
  m}_{S_2}\sim {\mathcal O}(1)$ GeV, the second generation light
sterile neutrino acts as the  mediator of dominant double beta decay
while the heavy quasi-degenerate pair of the first and the third
generation sterile neutrinos mediate resonant leptogenesis. 
Because of the resonant leptogenesis constraint, we note that either
Case (a) or Case (b) is permitted, but not both.   

\begin{table}
\begin{tabular}{|c|c|c|c|c|}
\hline
$m_{s_1}$&$m_{s_2}$&$m_{s_3}$&Baryon &$T_{1/2}^{0\nu}$\\
 (GeV)&(GeV)&(GeV)&asymmetry&$(10^{25}yrs.)$\\ \hline
 1 & 500 & 500 &$3.73\times10^{-10}$ & 2.72 \\ \hline
10 & 500 & 500 &$3.5\times10^{-10}$ & 16.01 \\ \hline
 500 & 1 & 500 &$4.2\times10^{-10}$ & 0.0494\\ \hline
 500 & 3 & 500 & $4.1\times10^{-10}$ & 2.19\\ \hline
\end{tabular}
\caption{Predictions for baryon asymmetry and double-beta decay
  half-life as a function of sterile neutrino masses.}
\label{bsymm}
\end{table}
 Our predictions for the double beta decay half-life and the baryon
 asymmetry in Case (a) and Case (b) are presented in Table
 .\ref{bsymm}. It is clear that for smaller mass eigen values of
 sterile neutrinos in Case (a) or Case (b), it is possible to saturate
 current experimental limit on the double-beta decay half-life while
 explaining the right order of magnitude of the baryon
 asymmetry. Thus, in addition to the Case (a) found in
 ref.\cite{bpn-mkp:2015}, we have shown another possible alternative scenario as
 Case (b).

Before concluding this section certain interesting results on thermal
leptogenesis derived earlier are noted. Thermal leptogenesis with a hybrid seesaw and
RH neutrino dark matter have been proposed by introducung additional
$U(1)$ gauge symmetry \cite{Deppisch:hybrid:2014}. Thermal leptogenesis  in extended seesaw models have been investigated earlier \cite{Hirsch:extended:lepto:2007,Sarkar:ge:2006,Ma-sahu-sarkar:2007} which are different from our cases reported here and earlier \cite{bpn-mkp:2015,bpn-mkp:disp:2015}. Possibilities of 
falsyfying high-scale leptogenesis on the basis of certain LHC results 
and also on the basis of LFV and $0\nu\beta\beta$ decay results have been suggested \cite{Deppisch:BAU-bb:2015} . Prospects of dark matter in the minimal
inverse seesaw model has been also investigated in ref.\cite{Abada:inv:DM:2014}

\section{ SUMMARY AND DISCUSSION}\label{sec.9}

Reviewing the contributions already made, we have discussed how the added
presence of singlet fermions which manifest as singlet neutrinos of
Majorana type, effectively cancels out the would-be dominant type-I seesaw contribution to
light neutrino masses. The neutrino masses consistent with the
 oscillation data are now governed by the classic inverse seesaw formula
(renamed as extended seesaw formula), or
type-II seesaw, or even linear or double seesaw under their respective
limiting conditions.  But the dominant part of the singlet neutrino
mass is given by the type-I
seesaw mechanism where the role of the Dirac neutrino mass matrix $M_D$ is
replaced by the $N-S$ mixing matrix $M$. The cancellation  mechanism
of type-I seesaw term for the active light neutrino masses is universal in the
sense that appropriate extensions of SM, left-right gauge theories,
Pati-Salam
Model, and SO(10), SUSY or non-SUSY can accommodate  it  leading to the dominance of seesaw mechanism of another type. In the
cases of extended  seesaw, linear seesaw, or
type-II seesaw dominance, the double beta decay rates in the $W_L-W_L$
channel are dominated by the exchange of light sterile neutrinos with 
 masses in the range of ${\cal O}  (1-10)$ GeV. With type-II dominance,
the second and third
generation sterile neutrinos could be heavy and quasi-degenerate, and explain baryon
asymmetry of the universe through resonant leptogenesis. The models
also predict a rich variety of results for LFV decays and
leptonic non-unitarity effects.
In SUSY GUTs in the absence of added fermion singlets baryogenesis via leptogenesis through decays of heavy RH
neutrinos is usually affected by gravitino problem. A possible
solution to this
 is the well known TeV scale resonant leptogenesis which requires extremely
small mass difference between the heavy quasi-degenerate pair leading to their
unity mass ratio upto very high degree of accuracy. The presence
of sterile neutrinos considerably alleviates this problem by changing this fine tuned mass ratio to a value as large as $\sim 10$. In SUSY SO(10) leptogenesis
under gravitino constraint is achieved even for large hierarchical masses
of RH neutrinos. When the model is extended by fermion singlets, the singlet
neutrinos which mix with RH neutrinos also acquire hierarchical masses
below $10^6$ GeV. The hierarchical singlet neutrinos decay through
their mixings with RH neutrinos to
generate the desired  leptonic CP-asymmetry. In these models also the
light neutrino mass formula is due to the extended seesaw ( or inverse
seesaw). One simple reason for the success of the sterile neutrino
assisted leptogenesis in SUSY SO(10) is the smallness of mixing angle $\xi$ between the lighter
sterile neutrino and the heavy RH neutrino with $\sin\xi \sim {M/M_N}
\sim 10^{-5} \to 10^{-6}$. This reduces the decay rate of the sterile
neutrino considerably to
satisfy the out-of-equilibrium condition which forms an important
ingredient for the generation of CP-asymmetry.   
\\

In the presence of sterile neutrinos, type-II seesaw dominance is achieved
with $U(1)_{B=L}$ breaking scale much lower than the mass of the Higgs
triplet $\Delta_L(3,1,-2,1)$. This mechanism makes it possible to
have type-II seesaw formula even with TeV scale of $W_R$ or $Z_R$ boson whereas in the conventional attempts in
SUSY or non-SUSY SO(10), or split-SUSY models, the type-II seesaw dominance required
these $W_R,Z_R$ boson masses to be near the GUT-Planck scale. In the
standard lore in the proposed models of
type-II seesaw dominance with very large $W_R$ and RH neutrino masses,
the non-standard contribution to double beta decay in the $W_L-W_L$
channel, damped by the heavy  $\Delta_L(3,1,-2,1)$ boson propagator, is
 negligible. But this concept is overthrown when type-II seesaw
dominance is assisted by sterile neutrinos. In the new scheme even
though the heavy  $\Delta_L(3,1,-2,1)$ boson exchange contribution is
negligible, the double beta decay mediation by the exchange of
light sterile neutrinos in the $W_L-W_L$ channel predicts dominant decay
rates saturating the current experimental limits. This new model of
type-II seesaw dominance predicts resonant leptogenesis in the  non-SUSY
SO(10) originating from the near TeV scale masses of quasi-degenerate
sterile neutrinos of the other two generations. 

While GUTs
like SUSY SO(10) have a rich structure of dark matter candidates, even
in the presence of sterile neutrinos, some aspects of embedding dark matter and
their detection possibilities in the SM extensions have been
discussed in
refs.\cite{Cheon:2007,Cheon:2009,Kang-Patra:2014}. The SUSY and non-SUSY SO(10)
GUTs considered under these seesaw mechanisms satisfy coupling unification and proton life time
constraints, the latter being accessible to ongoing search experiments 
\cite{mkp-sahoo:Proc:2014,mkp-sahoo:NP:2015,bpn-mkp:2015,
mkp-arc:2010,app:2013,pas:2014}.
    
In this review we have considered the class of models where heavy RH 
Majorana mass terms are present satisfying the conditions
$ M_N > M >>
M_D,\mu_S$ under which the generalised form of the neutral fermion
matrix gives different seesaw formulas. In these models the type-I
seesaw, if not cancelled by using the decoupling criteria and two-step
block diagonalization process, would have given dominant contributions. A common feature of all these
models are dominant double beta decay in the $W_L-W_L$ channel
mediated by light sterile neutrinos as well as leptogenesis generated
by heavier sterile neutrinos of the other two generations. 

The singlet neutrinos needed for these models are found to have mass
ranges between few GeV to $\sim 1 $ TeV. They are the mixed states of
added fermion singlets $S_i$ and heavy RH neutrinos $N_i$ where the
latter are 
in the spinorial representation of SO(10).  The mass terms of fermion
singlets violate the global lepton number symmetry of the SM. As such these
masses are required to be as light as possible according to 'tHooft's
naturalness criteria \cite{tHooft:1975}. In other words the global
lepton number symmetry protects these masses naturally and prevents
them from becoming superheavy. This is a special advantage in favour
of TeV scale seesaw mechanisms as well as the new type-II intermedate
-scale seesaw mechanism \cite{bpn-mkp:2015} due to the cancellation of the
type-I seesaw. Further we have brought down the $U(1)_R\times U(1)$
breaking scale in non-SUSY SO(10)  to be accessible to LHC and future
accelerator searches by a number of new physical processes including
the $Z'$ boson. The predicted proton lifetime has been noted to be
accessible to ongoing searches \cite{Proton-decay:expt}.

There are a
number of interesting models assisted by TeV scale pseudo Dirac neutrinos \cite{
mkp-sahoo:Proc:2014,
psb-rnm:2010,ap:2012,Das-Okada:2013,Kang:2016,Blanchet:2010} where such
heavy RH Majorana neutrino masses are either absent or, if present, they do not
satisfy the decoupling criteria. Details of phenomenology and predictions of 
such models  are beyond the scope of the present review. Likewise the
interesting possibilitis of detection of gauge singlet neutrinos through their
displaced vertices \cite{bpn-mkp:disp:2015,Helo,Antusch:2016} and the
renormalization group impacts \cite{RG-inv} in the presence of inverse
seesaw  have been excluded from  present discussions.

\vspace{2cm}
\section{ACKNOWLEDGMENT}
The authors are thankful to Professor R. N. Mohapatra and DR. Mainak
Chakraborty for discussions.
M. K. P. acknowledges research grant No.SB/S2/HEP-011/2013  from the
Department of Science and Technology, Govt. of
India. B. P. N. acknowledges a junior research fellowship from the
Siksha 'O' Anusandhan University, Bhubaneswar. 
\section{DECLARATION}
While submitting this manuscript for publication, the authors declare that they have no conflict of interest with any
individual or organisation, private or government, whatsoever.

\end{document}